\documentclass[acmsmall,screen,nonacm]{acmart}

\AtBeginDocument{%

  \providecommand\BibTeX{{%
    \normalfont B\kern-0.5em{\scshape i\kern-0.25em b}\kern-0.8em\TeX}}}

\setcopyright{acmcopyright}
\copyrightyear{2018}
\acmYear{2018}
\acmDOI{XXXXXXX.XXXXXXX}

\acmJournal{JACM}
\acmVolume{37}
\acmNumber{4}
\acmArticle{111}
\acmMonth{8}

\usepackage{booktabs}
\usepackage{multirow}
\usepackage{graphicx}
\usepackage{lscape}
\usepackage{makecell}
\usepackage{tabularray}
\usepackage{booktabs}
\usepackage{graphicx}
\usepackage{amsmath}

\usepackage{caption}
\usepackage{subfigure}
\usepackage{caption}
\usepackage{pifont}
\usepackage{enumitem}

\newcommand{\cmark}{\ding{51}}%
\newcommand{\xmark}{\ding{55}}%
\newcommand{\norm}[1]{\left\lVert#1\right\rVert}

\DeclareMathOperator*{\argmin}{arg\,min}
\begin{document}

\title{A Survey of Privacy Threats and Defense in Vertical Federated Learning: From Model Life Cycle Perspective
}


\author{Lei Yu}
\authornote{Both Lei Yu and Meng Han contributed equally to this research as the co-first author and co-corresponding author.}
\affiliation{%
  \institution{Rensselaer Polytechnic Institute}
  \city{Troy}
  \country{USA}}
\email{yul9@rpi.edu}

\author{Meng Han}
\authornotemark[1]
\email{mhan@zju.edu.cn}
\author{Yiming Li}
\email{li-ym@zju.edu.cn}
\author{Changting Lin}
\email{linchangting@zju.edu.cn}
\affiliation{%
  \institution{Zhejiang University}
      \city{Hang Zhou}
  \country{China}
}

\author{Yao Zhang}
\email{qianyao.zy@antgroup.com}
\author{Mingyang Zhang}
\email{zhangmingyang.zmy@antgroup.com}
\author{Yan Liu}
\email{bencao.ly@antgroup.com}
\author{Haiqin Weng}
\email{haiqin.wenghaiqin@antgroup.com}
\affiliation{%
  \institution{AntGroup}
    \city{Hang Zhou}
  \country{China}
  }
\author{Yuseok Jeon}
\email{ysjeon@unist.ac.kr}
\affiliation{
  \institution{Ulsan National Institute of Science and Technology}
    \city{Ulsan}
  \country{South Korea}
}

\author{Ka-Ho Chow}
\email{kachow@cs.hku.hk}
\affiliation{%
  \institution{The University of Hong Kong}
    \city{Hong Kong}
  \country{China}
  }

\author{Stacy Patterson}
\email{sep@cs.rpi.edu}
\affiliation{%
  \institution{Rensselaer Polytechnic Institute}
    \city{Troy}
  \country{USA}
  }
\renewcommand{\shortauthors}{Yu and Han, et al.}

\begin{abstract}
Vertical Federated Learning (VFL) is a federated learning paradigm where multiple participants, who share the same set of samples but hold different features, jointly train machine learning models.  Although VFL enables collaborative machine learning without sharing raw data, it is still susceptible to various privacy threats. In this paper, we conduct the first comprehensive survey of the state-of-the-art in privacy attacks and defenses in VFL. We provide taxonomies for both attacks and defenses, based on their characterizations, and discuss open challenges and future research directions. Specifically, our discussion is structured around the model's life cycle, by delving into the privacy threats encountered during different stages of machine learning and their corresponding countermeasures. This survey not only serves as a resource for the research community but also offers clear guidance and actionable insights for practitioners to safeguard data privacy throughout the model's life cycle.

\end{abstract}




\maketitle

\section{Introduction}\label{Intro}
Federated Learning (FL)~\cite{kairouz2021advances} has emerged as a revolutionary paradigm for collaborative machine learning, allowing multiple entities to jointly train a global model without sharing raw data. It was first proposed by Google in 2016~\cite{mcmahan2016federated}, which enables cross-device model training for keyboard input prediction without exchanging local user data.
Compared to conventional machine learning, the benefits of FL are manifold. It ensures data privacy and security by eliminating the need to centralize data, which facilitates adherence to data protection regulations such as the European Union's General Data Protection Regulation (GDPR) \cite{GDPR} and the United States' California Consumer Privacy Act (CCPA) \cite{CCPA}. On the other hand, FL improves scalability, reduces the costs associated with data transfer and storage, and allows real-time model updates by utilizing data directly at its source. Comprehensive surveys of the concepts and applications of FL are available in the literature \cite{Yang2019fedsurvey,Li2020fedsurvey,Li2023fedsurvey}.

Generally, FL can be divided into two types~\cite{Yang2019fedsurvey}: horizontal federated learning (HFL) and vertical federated learning (VFL), according to how data is partitioned in the sample and feature space. Both HFL and VFL train a global model without sharing raw data among participants. HFL requires all participants to possess the same feature space but different samples. In VFL, all participants share the same samples but hold different features. In addition, FL is commonly implemented in two scenarios: cross-device and cross-silo, as outlined in \cite{kairouz2021advances}. Cross-device FL involves a large number of mobile devices as participants, whereas cross-silo FL encompasses a much smaller and more controlled group of organizations. HFL can be either cross-device or cross-silo FL, whereas VFL is typically cross-silo.

In recent years, VFL stands out as a promising approach that enables the collaboration of parties possessing complementary information about the same set of samples. This powerful technique has found wide applications in various domains, ranging from healthcare~\cite{hu2022vertical} and finance~\cite{webank2020vfl} to smart cities and Internet of Things (IoT)~\cite{teimoori2022secure} scenarios. However, as the applications of VFL accelerate, so do concerns about the security and privacy risks it poses. Understanding privacy challenges and formulating effective defense strategies is of paramount importance. VFL, despite its distributed and privacy-preserving nature, is susceptible to various privacy attacks. These attacks exploit vulnerabilities inherent in the collaborative learning process to infer sensitive information from the model output or its updates, thereby undermining data privacy and confidentiality.

In response to these growing concerns, researchers and practitioners have made significant efforts in exploring both privacy vulnerabilities and defense mechanisms in VFL. However, there is a lack of a comprehensive survey dedicated exclusively to the privacy aspects of VFL. Existing surveys on federated learning privacy~\cite{MOTHUKURI2021619,TRUONG2021102402,lyu2020threats,chen2022federated} predominantly address HFL. While the survey by Yin et al.~\cite{yin2021comprehensive} covers VFL, it does not offer a distinct analysis of VFL privacy risks separate from those in HFL, and it does not delve into detailed discussions on the privacy attacks and defense mechanisms pertinent to VFL. Moreover, although there are a few extensive surveys on the broader advancements of VFL~\cite{liu2023vertical,yang2023survey,khan2022vertical,wei2022vertical}, they do not concentrate solely on privacy issues. Consequently, these works do not provide an in-depth examination of VFL's privacy landscape, nor do they offer a detailed taxonomy and analysis of privacy threats and defenses.
Given the growing importance of VFL, the district privacy challenges it presents, notably inadequate literature coverage and rapid advancement in VFL privacy research, a survey dedicated to VFL privacy is necessary.

Therefore, this survey paper aims to fill the gap by providing a comprehensive overview of state-of-the-art work in the field, offering a systematic analysis of the various privacy threats and the defense strategies proposed to safeguard sensitive data in VFL. In this paper, we discuss the distinction of VFL privacy from those in HFL, and provide a comprehensive framework for understanding the threat model and defense strategies. We meticulously examine a wide spectrum of privacy attacks targeting VFL systems, including label inference attacks, model extraction attacks, and feature inference attacks. We also present a detailed survey of defense strategies against these attacks, such as differential privacy, homomorphic encryption, and random perturbation. Compared to existing literature surveys on VFL, our work is concentrated exclusively on VFL privacy attacks and defenses, differentiates them between machine learning stages, and provides a more granular level of detail and a more specific taxonomy.

\begin{figure}[H]
\includegraphics[width=1\textwidth]{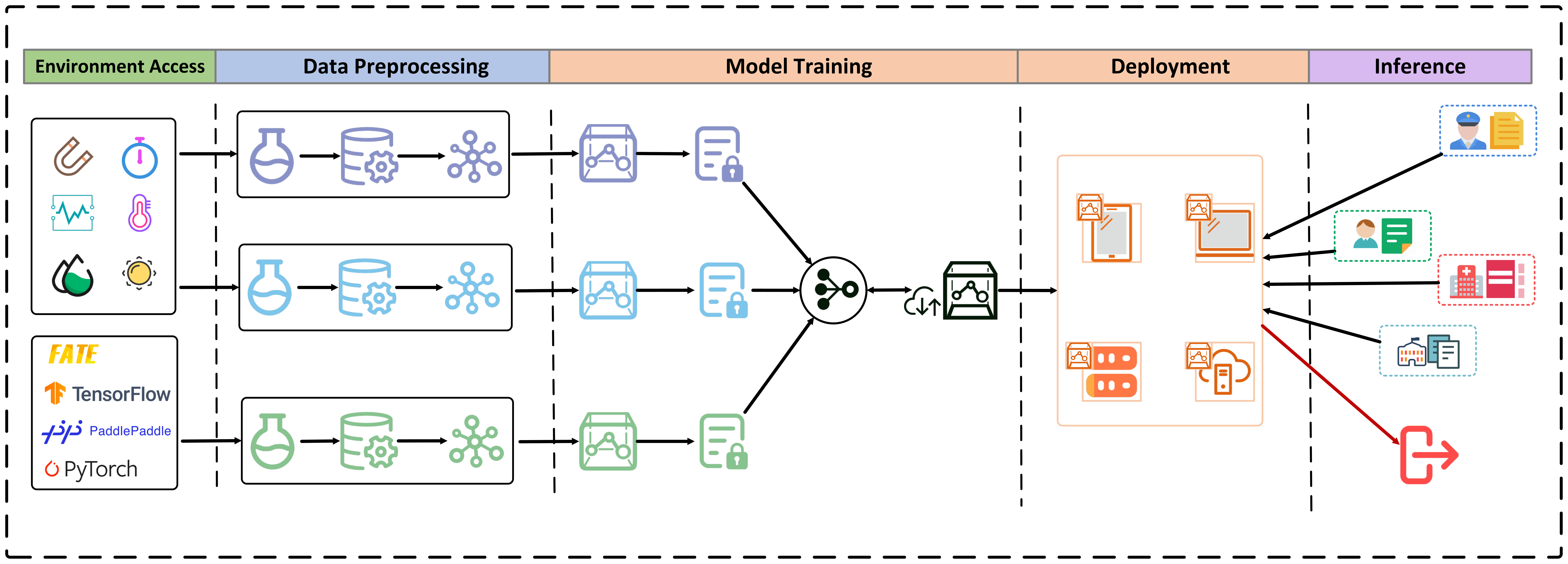}
\caption{Different Phases of Machine Learning Life-cycle}\label{fig:VFLoverview}
\centering
\end{figure}

In particular, our survey is structured around the model’s life cycle. As shown in Fig. \ref{fig:VFLoverview}, a typical machine learning life cycle includes environment access, data preprocessing, model training, model deployment and model inference. Environment access is the initial setup phase of machine learning, which typically involves data access, data collection, and setup of computational resources. Data preprocessing involves operations such as data normalization, data cleaning, feature selection, and data partitioning. In VFL, an additional critical step in data preprocessing is entity alignment, which ensures that data from different sources but pertaining to the same entity are correctly matched. Our paper will elaborate on this process as well as the rest of the phases in VFL.
We discuss the privacy threats that arise during these different phases and the respective countermeasures. By doing so, we aim for this survey to provide a clear characterization and differentiation of privacy attacks and defenses across various phases, while offering practical guidance for practitioners, enabling them to protect data privacy effectively throughout the entire life cycle of the model. Finally, we identify the research gaps, crucial challenges and areas for future research in VFL privacy. Our objective is to inspire the development of novel and robust privacy-preserving solutions that will boost the widespread adoption of VFL in real-world applications.

In summary, the main contributions of this work are as follows:
\begin{itemize}
    \item We provide an exhaustive examination of state-of-the-art works on privacy attacks and defenses in VFL from model life cycle perspective, facilitating a comprehensive understanding of the current landscape across different VFL phases. 
    \item We provide a comprehensive overview of key concepts of VFL and privacy techniques. Additionally, we characterize the difference in privacy threats between HFL and VFL, showing the importance of specialized treatment for privacy concerns in VFL.
    \item We propose an extensive taxonomy to systematically characterize privacy threats and defenses in VFL, according to their assumptions, application scenarios, and strategies.
    \item We provide in-depth discussions on the key challenges of privacy issues of VFL, open research problems, and future research directions.
\end{itemize}

\section{Background}\label{Backg}
Federated learning (FL) is a distributed machine learning paradigm that allows multiple participants to collaboratively train a global model by following a specified training protocol. Each participant provides a set of private training data, which remains stored on their local devices and is not directly shared during training. The way in which data is distributed among participants determines the type of FL adopted. In this section, we present two prominent types of FL: HFL and VFL. We provide an overview of the VFL training protocol, highlight the differences between HFL and VFL and discuss the connection between VFL and split learning.    

\subsection{Types of Federated Learning}
Federated Learning can be categorized into two primary types based on the partitioning of data in the sample and feature spaces: Horizontal Federated Learning (HFL) and Vertical Federated Learning (VFL) \cite{yang2019federated}.

\noindent\textbf{HFL vs. VFL on Data Partition:}
\begin{itemize}
    \item Horizontal Federated Learning (HFL): In HFL, participants have data with the same feature space but different samples. This is common when different entities collect similar types of data on different sets of users. For example, two hospitals might perform identical medical tests (features) but for different patients (samples).
    \item Vertical Federated Learning (VFL): VFL is used when participants have data on the same samples but with different features. This often occurs when different entities have different types of data on the same set of users. For example, a bank might have financial transaction data (a specific feature set), and an e-commerce platform could hold the same individual's purchase data (a different feature set).
\end{itemize}
There exist additional types of FL such as Federated Transfer Learning (FTL). FTL comes into play when participants have both different feature spaces and different samples. FTL involves transferring knowledge from one domain to another. For readers who are interested, the VFL survey by Liu et al. ~\cite{liu2023vertical} offers a comprehensive overview on different types of FL.

While both HFL and VFL aim to train a global model without sharing raw data, they differ in several key aspects:

\noindent\textbf{HFL vs. VFL on Training Process:}
\begin{itemize}
    \item HFL: In each training round, each participant trains a local model on their dataset using a shared global model as the starting point. A server aggregates the model updates (i.e., parameters or gradients) from all participants to update the global model. This process is repeated until the global model converges.
    \item VFL: Given that participants have different features for the same samples, the training process of VFL requires additional data processing step, i.e., entity alignment, which align the sample IDs to ensure that they are working on the same set of samples. Each participant collaboratively train a unified global model by exchanging intermediate model outputs and gradients for the shared samples. Each participant's local model serves as a segment or sub-model of the global model.
\end{itemize}

\noindent\textbf{HFL vs. VFL on Inference Process:}
\begin{itemize}
    \item HFL: The output of the HFL training process is a global model shared among all participants. Once they receive this model, each participant can independently perform inferences using its local model copy.   
    \item VFL: The output of VFL is essentially a composite of local sub-models from each participant. Each participant's local model only focuses on their specific feature set. They must collaborate to make inferences.
\end{itemize}

\noindent\textbf{HFL vs. VFL on Model Architecture:}
\begin{itemize}
    \item HFL: An identical or similar model architecture is used across all participants.   
    \item VFL: Different model architectures may be employed by participants for their local sub-models, tailored to their specific feature sets.
\end{itemize}

These differences lead to significant distinctions in privacy considerations between VFL and HFL, which we will explore in detail in the following section.

\subsection{Privacy Considerations on HFL v.s VFL}
\label{ssec:privacyvfl}
For HFL, the main privacy concern is that the model updates may cause the leakage of information about individual samples of a participant.
In contrast, VFL presents more intricate privacy concerns, stemming from the unique differences in its training, inference procedures, and local sub-model architectures. Generally, VFL is more susceptible to privacy breaches, compared with HFL, due to the following reasons:
\begin{itemize}[leftmargin=*]
    \item \textbf{Exchange of direct data representation}:  In VFL, the gradients and intermediate model outputs often represent one-pass computations on or transformations of a specific set of local samples. As a result, they are more closely tied to the raw data compared to model updates in HFL, making them more susceptible to reverse engineering to infer the original data. In contrast, HFL's model updates are typically aggregated over multiple iterations and many samples, making them more generalized and less revealing.
    \item \textbf{Frequency of direct data representation exchange}: VFL requires participants to collaborate and exchange intermediate output for each iteration during the training process and for each inference procedure. Compared with HFL, this increased message exchange can provide more opportunities for adversaries to collect and analyze the data.
    \item \textbf{Aligned sample computation}: In VFL, the necessity for sample alignment during training means that an insider adversary can readily correlate the intermediate outputs or gradients with the specific samples they are based on in each training iteration. This makes those samples more vulnerable to targeted attacks. Conversely, in HFL, an adversarial participant only gets access to a model that has been averaged across several participants. This aggregation provides a layer of obfuscation, making it more challenging for an adversary to target data from a specific participant.
    \item \textbf{Feature correlation complexity}: Given that exchanged outputs are derived from the same set of samples in VFL, they might inadvertently reveal relationships between private features owned by different participants for the same sample, leading to unintended data disclosures. In contrast, samples from different participants in HFL are typically independent of each other.
\end{itemize}
Given the distinct privacy challenges in VFL, it is essential to deepen our understanding of its potential privacy threats. This becomes even more critical considering that research on privacy in VFL lags far behind that of HFL.

\subsection{VFL Framework}
In this section, we present an overview of the VFL framework from aspects of architectures, the protocol participants, the general process and connection with split learning.

\subsubsection{VFL Architectures}
VFL can be divided into two types of architectures: aggregate VFL (\emph{AggVFL}) and split VFL (\emph{SplitVFL}) according to whether the local model outputs of all parties are simply aggregated or fused through a trainable model to produce the final prediction~\cite{liu2023vertical}.
Suppose that each VFL participant $p_i$ has its local model $M_i$ parameterized by $\theta_i$ and computes the output with its local data feature $x_i$ of sample $x$ by $M_i(x_i,\theta_i)$. The final VFL model $M$'s output on sample $x=(x_1,x_2,\ldots,x_N)$ is computed over all the local model outputs from all $N$ participants.
\begin{equation}
    M(x) = F(M_1(x_1,\theta_1), M_2(x_2,\theta_2), \ldots, M_N(x_N,\theta_N))
\end{equation}
where $F$ is the final prediction function.
Accordingly, VFL is
\begin{itemize}
    \item Aggregate VFL (\emph{AggVFL}), if $F$ simply serves an aggregate function that is non-trainable, such as sigmoid (for logistic regression) or an optimal split finding function (for tree) that aggregates intermediate results from participants.
    \item Split VFL (\emph{SplitVFL}), if $F$ is a sub-model with trainable parameters, such as the final fully connected layer in Split learning~\cite{thapa2022splitfed}.
\end{itemize}
The function $F$ can be implemented either on a central server or as part of a participant which we refer to as an active party, explained in the following section. 

\textbf{VFL v.s. Split Learning}\quad VFL is designed for scenarios where datasets across different participants share the same samples but different features. On the other hand, Split Learning~\cite{gupta2018distributed}  is a distributed learning approach in which the neural network model itself is divided among participants. SplitVFL can be understood as the application of split learning techniques to vertically partitioned data.  Therefore, our survey includes privacy attacks and defenses in the context of split learning that are also relevant to SplitVFL.

\subsubsection{VFL Participants}
In VFL, participants typically play three types of roles based on the nature of their data and their involvement in the training process: active parties, passive parties, and a coordinator.
\begin{itemize}[leftmargin=*]
\item \textbf{Active Parties}: An active party has the labels of the samples. It either solely has the labels of a sample set or has both data features and labels. The local model of the active party is typically the final sub-model responsible for label prediction in the global VFL model, often referred to as \textbf{top models} in SplitVFL. In the training process, the active party calculates the training loss and drives gradient descent. 
\item \textbf{Passive Parties}: These participants have features of sample sets without labels, and there are typically multiple passive parties. Their primary role involves computing and transmitting intermediate model outputs. In return, they receive gradients that they use to update their local models. These local models are often referred to as \textbf{bottom models} in SplitVFL which take local data samples as input.
\item \textbf{Coordinator Server}: A coordinator server is responsible for coordinating the communication among participants and handling the aggregation of updates. It also plays a crucial role in the security of VFL by managing cryptographic operations, including key distribution and data encryption and decryption. The coordinator can be an external trusted third party, though in many scenarios discussed in this paper an active party assumes this role.
\end{itemize}

\subsubsection{General VFL Procedure}
A general VFL  procedure consists of three key steps, i.e., entity alignment, training, and inference.
\begin{itemize}
    \item Entity alignment step: Before the training step, VFL participants need to find the identifiers of common samples. This process, called entity alignment, uses a private set intersection, which can avoid revealing the sample IDs that are not in the intersection set of participants. We will discuss these techniques in detail in Section \ref{sec:preprocess}. 
    \item VFL training step: After entity alignment, participants collaboratively train a global VFL model with aligned samples. This training, particularly for neural networks, typically distributes forward and backward propagation processes across all involved parties. During forward propagation, each party computes its local model output on a mini-batch of aligned samples. The computed outputs, referred to as intermediate results, are then forwarded to the active party. The active party uses the intermediate results from the passive parties to evaluate the loss function on its top model with ground-truth labels. During backward propagation, the active party computes the gradients of the loss function with respect to both its top model and the received intermediate results. It then sends the gradients of the intermediate results back to passive parties. The passive parties use these gradients to compute the gradients relative to their own model parameters. Each party uses gradient descent to update its local model. 
    For a detailed explanation, let's examine a typical two-party VFL process:
\begin{itemize}
    \item $X_A, X_B$: Input features from active party $A$ and passive party $B$.
    \item $f_A, g_B$: Local models of party $A$ and $B$.
    \item $\theta_A, \theta_B$: Model parameters of $f_A$ and $g_B$.
    \item $Y_A$: Labels held by active party $A$.
\end{itemize}
\begin{enumerate}
    \item[+] \textbf{Forward Propagation:}
    \begin{itemize}
        \item $B$ computes intermediate result: $R_B = g_B(X_B; \theta_B)$ and forwards $R_B$ to $A$.
        \item $A$ computes the loss: $L = \text{Loss}(f_A(X_A, R_B; \theta_A), Y_A)$.
    \end{itemize}
    
    \item[+] \textbf{Backward Propagation:}
    \begin{itemize}
        \item $A$ calculates the gradients $\nabla_{\theta_A} L$ and sends $\nabla_{R_B} L$ to $B$.
        \item $B$ computes its gradients: $\nabla_{\theta_B} L = \nabla_{\theta_B}R_B\times \nabla_{R_B} L$.
    \end{itemize}
    
    \item[+] \textbf{Parameter Update:}
    \begin{itemize}
        \item Both $A$ and $B$ update their parameters ($\theta_A, \theta_B$) using the gradients to minimize $L$.
    \end{itemize}
\end{enumerate}

\item VFL inference step: After a VFL model is trained, all parties collaboratively perform the inference. An inference request for a sample requires the VFL service coordinator to send the corresponding sample ID to all parties.
\end{itemize}

\section{Threat Model}
In this section, we systematize privacy threats in VFL according to the attacker's goal, their knowledge about the target system, their intent, and their capability. It is important to note that these aspects can vary at different stages of the machine learning life cycle, leading to various types of privacy threats. 
This survey specifically addresses threats posed by insider adversaries who are participants within the VFL protocol. It is because the threats posed by external adversaries are analogous to those encountered in traditional distributed systems and centrailized ML systems and they do not exhibit distinct characteristics unique to VFL.

\subsection{Attacker's Goals}
\label{ssec:attackgoal}
The goal of a privacy attack can be defined in terms of the type of private information targeted by the adversary. In VFL, the attacker could target the label data, training data and the model of a participant, and the features of the samples to be inferred. Based on these goals, we classify privacy attacks into feature inference attacks, label inference attacks, and model extraction attacks, discussed in the following. In this paper, the term "data inference attack" is used to encompass both feature inference attacks and label inference attacks.

\noindent\textbf{Feature Inference Attack (FIA)}\quad
In VFL the features of a sample are often collected by participants and distributed among multiple parties as private assets. It is often referred to as a data/input reconstruction attack. 
In a feature inference attack, the adversary's objective is to extract private features of data samples, which could be either training data or inference data. Typically, the adversary is the active party in VFL and targets the passive parties to obtain their private data. 

\noindent\textbf{Label Inference Attack (LIA)}\quad
In VFL, the label data of the active party is considered a valuable and sensitive asset. For a label inference attack, the victim is typically the active party because it has label data,  and the adversary is a passive party who attempts to obtain the labels of samples to which he has access.  

\noindent\textbf{Model Extraction Attack (MEA)}\quad
VFL has a model split between multiple participants. For a model extraction attack, the adversary can be any of the parties and tries to extract the model part from another party. Therefore, we will discuss these attacks in both directions: from the active party to a passive party and vice versa. It is worth noting that model extraction is actually achieved by training a surrogate model that is close to the target victim model. However, achieving exact functional equivalence is difficult in practical applications, and most existing practical ME attacks focus on achieving high accuracy with respect to prediction correctness and fidelity with respect to a decision boundary similar to the victim model.

We note that, in the context of HFL, there are other types of privacy attack with regard to different goals, including the membership inference attack and the property inference attack. \emph{Membership inference attack}~\cite{nasr2018comprehensive,melis2019exploiting,shokri2017membership} aims to determine whether a sample is part of the training dataset.  However, it does not apply to VFL because VFL requires entity alignment among different parties that inevitably reveals the sample membership information. \emph{Property inference attack}~\cite{ganju2018pia,kerkouche2023client} is a privacy violation in federated learning where an adversary attempts to infer certain higher-level properties or patterns of the training data owned by individual clients. To the best of our knowledge, there has been little research on this particular attack in the context of VFL.

\subsection{Attacker's Knowledge}
Generally, an attacker can have varying degrees of knowledge about the target model and VFL system during the ML life cycle, resulting in the use of various attack techniques and effectiveness. The knowledge an adversary can access and use for attacks depends on the stages in ML life cyle, its role in VFL (active party, passive party or coordinator), the availability and types of auxiliary data and cryptographic protocols used (such as homomorphic encryption). We classify it into the following categories, based on the type of knowledge, with two-letter abbreviations. For insider attackers, the types of knowledge in the following are meant to specify additional information that goes beyond the locally accessible samples. By default, an insider attacker can access their own local data and model parameters, except in cases where these are encrypted to prevent local access.

\noindent\textbf{Data Characteristics (K-C)}\quad
An adversary may have information about the characteristics of training or inference data, such as population distribution, class imbalance, and whether the features are binary or numerical. The adversary can use this information to devise effective data inference attacks.

\noindent\textbf{Auxiliary Samples (K-A)}\quad
Many attacks proposed in VFL assume that the adversary has access to an auxiliary labeled dataset of limited size from the same domain of the training and testing dataset.
By using the auxiliary data, the adversary can train a surrogate model that can serve for FIA, LIA and MEA attacks. This type of knowledge could be available not only during the training time but also all the other stages of ML life cycle, depending on if the adversary has any chance to access the data from the same domain. They may obtain the auxiliary data from public sources.


\noindent\textbf{Gradients (K-G)}\quad
During training time, participants in the VFL exchange gradients to update their local models. An adversary who is a partcipant in the system can harness gradient information received from other parties to infer private data.

\noindent\textbf{Intermediate Results (K-I)}\quad
During both the training and the inference time of VFL, a passive party needs to forward its local model's output to another party (either an active party or another passive party) for the loss computation or model inference. The transmitted local model's output is called intermediate results. For an adversary who is the receiving party, he can use intermediate results to infer private data and model of the sending party.     

\noindent\textbf{Prediction Confidence Scores (K-S)}\quad
During the inference time, the VFL system can respond to a prediction request with confidence score that is a probability vector of predicted classes. An adversary can use the confidence score to infer private input data and extract models.

\noindent\textbf{Model Parameters (K-M)}\quad
An adversary may gain access to the global model or models of other participants during the training phase or after training. An insider attacker often has direct access to their local model parameters, unless otherwise specified.

\noindent\textbf{Model Hyperparameters (K-H)}\quad
An adversary may have information about the architecture of the target participant's local model. The adversary can use this information in different types of inference attack, i.e. train a clone model with the same architecture to infer the target participant's private data and model.

\subsubsection*{Attack Type based on Knowledge level.}
Based on the different levels of knowledge and access to the target system, attacks can be categorized into three types: white-box, black-box and gray-box attack.  

\noindent\textbf{White-box Attack (W-A)}\quad In a white-box attack, the adversary has complete knowledge of the internal details of the targeted VFL system or the participant. This includes knowledge of model architecture and parameters of the target participant.

\noindent\textbf{Black-box Attack (B-A)}\quad In a black-box attack, the adversary has limited or no knowledge of the internal details of the target. They typically can only query the target model and obtain the model output (i.e., either the final prediction scores or the intermediate results from the target participant). However, compared to traditional machine learning systems and HFL, black-box attacks in VFL may have unique restrictions on input access and advantages in information exchange. The primary restriction is that the adversary cannot access and control the input features from the target participant and therefore cannot determine the complete input for a query. On the other hand, an advantage is the potential access to gradients and intermediate results exchanged with the target participant.

\noindent\textbf{Gray-box Attack (G-A)}\quad A gray-box attack falls somewhere between white-box and black-box attacks in terms of knowledge and access. In this type of attack, the adversary leverages the partial information they have to devise their attacks. This can include the model architecture, some knowledge of local data or other system information of the target participant.

Because we focus on VFL and insider adversaries, the above attack types can exhibit subtle differences compared to traditional ML systems. All three types of attacks may have access to gradients exchanged during training, intermediate results from the target participant, and final prediction scores. In a black-box attack, it is common for the adversary to require auxiliary information, including K-C and K-A. However, existing white-box and gray-box attacks rarely use this auxiliary information, but are capable of leveraging them. Table \ref{tab:my_label} shows the correlation between the attack types and the knowledge types. 

\begin{table}
\caption{Attack type v.s. Knowledge. \cmark indicates that existing attacks of the specified attack type utilize the corresponding knowledge type, \xmark denotes no access to that knowledge type and - suggests that while no existing works have employed the knowledge type for that attack, it remains a feasible approach.}
\label{tab:my_label}
    \centering
\small
\begin{tabular}{|c|c|c|c|c|c|c|c|}
\hline
 & K-C &K-A & K-G & K-I & K-S & K-M & K-H  \\
\hline
White-Box  & - & - & \cmark & \cmark & \cmark & \cmark & \cmark \\
\hline
Black-Box & \cmark & \cmark & \cmark & \cmark & \cmark & \xmark  & \xmark \\
\hline
Gray-Box & - & - & \cmark & \cmark & \cmark & \xmark & \cmark \\
\hline
\end{tabular}
\end{table}

\subsection{Attacker's Behavior and Capability}
 There are the two primary types of attackers in VFL protocol based on their behavior:
\begin{itemize}
    \item Semi-Honest (or Passive) Attacker: A semi-honest participant is one who follows the protocol but may try to gain private information from the exchanged messages and computation results. They are passive adversaries who do not deviate from the VFL protocol. 
    \item Malicious (or Active) Attacker: A malicious participant is actively adversarial and may deviate from the protocol in an arbitrary way to learn sensitive information. They can send false information, manipulate learning algorithms, or attempt to sabotage the protocol's security.
\end{itemize}
Additionally, if there are multiple attackers, the attackers can be either \emph{non-colluding} or \emph{colluding}. Colluding attackers coordinate with each other to achieve a malicious goal.   

The capability of a malicious attacker defines his ability to manipulate the resources at his disposal, including the data, the algorithm, and the model.

\noindent\textbf{Input-Output Manipulation (IO-M)}\quad
An adversary can modify input features or intermediate results in VFL, termed data manipulation, which allows him to influence the behavior of the learning process and learn sensitive information. Data manipulation is applicable to both local data and communication between participants. 

\noindent\textbf{Learning Manipulation (L-M)}\quad
An adversary in VFL can alter its local learning algorithm and the goal and transmit malicious gradient updates generated from this altered learning process to other participants during training. By manipulating the learning process in this way, the adversary can guide the target participant's local model to release information that is more valuable for data inference attack.

\noindent\textbf{Gradient Manipulation (G-M)}\quad
An adversary can directly manipulate the gradients and transmit them to other participants during training. The key distinction from learning manipulation is that, in this case, the adversary does not change the learning process itself; instead, they only modify the gradients produced from the normal learning process.

Typically, an active attacker must ensure that the VFL system still retains normal functionality and high prediction accuracy under these manipulations, so that they can maintain the stealthiness of their attacks and successfully pass any validation without arousing suspicion.

\subsection{Attacker's Strategy}
The attack strategy describes how an adversary utilizes its knowledge and capabilities to execute feature inference attacks, label inference attacks, and model extraction attacks.

\noindent\textbf{Gradient Inversion (G-S)}\quad
As an insider participant, the adversary can exploit gradients received from other participants to infer private data and model information. When the target is the active party, the exchanged gradients might directly reveal the label information.  The adversary's key task involves deducing the mapping between the received gradients and the target's private information. This mapping can be established through different methods, including systems of equations, clustering, and machine learning models. Discovering and reversing the mapping is crucial for conducting effective inference attacks.

\noindent\textbf{Intermediate Result Inversion (I-S)}\quad
Similar to gradient inversion, intermediate results forwarded from the target party are also often used to reconstruct the target party's private data.

\noindent\textbf{Prediction Inversion (P-S)}\quad
The confidence scores generated from model predictions can be also leveraged to reconstruct the model's input. This process can be approached in various ways, such as model inversion attack~\cite{fredrikson2015model}, solving a system of equations, or training a generative model. 

\noindent\textbf{Surrogate Model (S-S)}\quad
Training a surrogate model is a common approach employed by adversaries in VFL for the purposes of data inference and model extraction. The adversary can use auxiliary samples or/and the information received during the training or inference such as gradients and intermediate results, to train a surrogate model that closely emulates the functionality of the target model. This approach can be used directly for model extraction. In addition, it can be used in label inference attacks by training a surrogate model to emulate the label prediction function. It can also be applied in feature inference attacks by using model inversion attack on the surrogate model that replicates the passive party's local model.

\section{Defense Framework}
In this section, we present a defense framework and main strategies to defend against privacy attacks described in Section~\ref{ssec:attackgoal}. Accordingly, the defense goals are to protect the privacy of the labels and the features, and the privacy of the model.

\subsection{Defense Capability}
The defense mechanisms for VFL privacy proposed in existing works can vary in terms of their requirements and capabilities. 
We identify the following different capabilities that may be required by various defense mechanisms in VFL:
\begin{itemize}
    \item Requiring a trusted third party: Many defense mechanisms assume the presence of a trusted third party that can perform sensitive tasks such as key management and gradient aggregation.
    \item Requiring control of the training procedure: Some defense mechanisms may require the ability to control the training procedure, such as gradient perturbation, differentially private training, and homomorphic encrypted training.
    \item Requiring participant coordination: Some defense methods rely on the cooperation of participants such as Multi-Party Computation (MPC).
    \item Requiring specialized security hardware: A few defense mechanisms utilize specialized security hardware like Trusted Execution Environments (TEEs) such as Intel-SGX and AMD enclave, to process data and execute code in a trusted isolated private enclave.
\end{itemize}

\subsection{Defense Strategy}
The strategies used by VFL privacy defense mechanisms can be broadly categorized into two types: cryptographic defense and non-cryptographic defense, depending on whether a mechanism relies on cryptographic primitives to protect privacy. 
We provide an overview of the representative methodologies used for each type.
\vspace{5pt}
\subsubsection{Cryptographic Defense Mechanisms}~

\noindent\textbf{Cryptographic Primitives:} we outline the common cryptographic techniques employed by cryptographic defense mechanisms as follows:
\begin{itemize}
    \item \noindent\textbf{Homomorphic Encryption (HE)}\quad
HE~\cite{rivest1978data} is adopted to allow computations to be performed on encrypted data without decrypting it first. The results of such computations remain encrypted and can only be decrypted by the owner of the private key. HE methods can be divided into partially homomorphic encryption (PHE) and fully homomorphic encryption (FHE). PHE supports limited computations, while FHE can support arbitrary computation on cipertexts. PHE schemes are used more widely in practice than FHE, such as Paillier~\cite{paillier1999public} and El Gamal~\cite{elgamal1985public}, due to their lower computation cost and faster speed. In this paper, homomorphic encryption is denoted as [[·]].

    \item \noindent\textbf{Functional Encryption (FE)}\quad
In FE~\cite{boneh2011functional}, decryption keys are associated with specific functions. It allows computing a specific function over a set of ciphertexts without revealing the inputs. It is generalization of publick-jey encryption. It uses functional decryption key $sk_f$ derived from the function $f$ and public key $pk$ to decrypt a ciphertext $c=Encrypt(pk,m)$, which is produced as a result $f(m)$. The decryption only reveals partial information $f(m)$ about the message $m$.

    \item \noindent\textbf{Secure Multi-Party Computation (MPC)}\quad
MPC~\cite{yao1982protocols} enables multiple parties to jointly compute a function over their inputs while keeping those inputs private. Three MPC techniques have been used for VFL:
\begin{itemize}
    \item\textbf{Secret Sharing (SS)}: Generally, a $(t,n)$-threshold SS~\cite{shamir1979share} is a cryptographic scheme that divides a secret among $n$ parties, where the secret can only be reconstructed when $m$ shares from an arbitrary subset of $m$ parties with $m \ge t$ are combined. Additive sharing~\cite{demmler2015aby} is a commonly used scheme in VFL. To additively share an $l$-bit value $a$ between two parties $P_0$ and $P_1$, the shares $a_0$ and $a_1$ are sent to $P_0$ and $P_1$ respectively and $a = a_0 + a_1 \mod 2^l$. We denote the shares $a_0 = \langle a \rangle_0^A$ and $a_1 = \langle a \rangle_1^A$.  Given two shared values $\langle a\rangle$ and $\langle b\rangle$,  for their addition $c =a+b$, each party $P_i$ can compute additive shares $\langle c\rangle_i$ = $\langle a\rangle_i$ + $\langle b\rangle_i$; for their multiplication $c=ab$, Beaver’s precomputed multiplication triplet technique~\cite{beaver1992efficient} is commonly used. Assume that $P_0$ and $P_1$ already share $\langle x\rangle$, $\langle y\rangle$ and $\langle z\rangle$ where $x$ and $y$ are random values and $z=xy$. $P_i$ computes $\langle e\rangle_i =  \langle a\rangle_i -  \langle x\rangle_i$ and $\langle f\rangle_i =  \langle b\rangle_i -  \langle y\rangle_i$. After reconstructing $e$ and $f$, $P_i$ computes $\langle c\rangle_i = i\cdot e \cdot f + f \cdot \langle x\rangle_i + e \cdot \langle y\rangle_i + \langle z\rangle_i$.
    
    \item \textbf{Oblivious Transfer (OT)}: OT~\cite{rabin2005exchange} is a protocol that allows the sender to send one of multiple values to the receiver without revealing any information about the other values, while the receiver can choose and obtain one value without revealing their selection information. In 1-out-of-2 OT~\cite{demmler2015aby}, the sender inputs two $l$-bit strings $s_0$ and $s_1$ and the receiver inputs a bit $c \in \{0,1\}$ and obliviously obtain $s_c$. The sender learns no information about $c$ and the receiver learns no information about $s_{1-c}$.  
    \item \textbf{Garbled Circuit (GC)}: GC~\cite{goldreich2003cryptography} enables secure two-party computation. It allows two parties to jointly evaluate a function (represented as a circuit) over their private inputs while keeping those inputs secret. Yao's Garbled Circuit protocol~\cite{yao1982protocols} and Goldreich-Micali-Wigderson (GMW) ~\cite{goldreich2019play} are two widely used techniques. The basic idea of Yao's GC is to let one party, called creator, encrypts the function $f$ to be computed and sends it along with its encrypted input $x$ to the other party called evaluator. The evaluator uses 1-out-of-2 OT to obtain its encrypted input $y$, and then evaluates $f$ on $x$ and $y$ to obtain the garbled output $z$. The creator finally provides a mapping from the garbled output to the plain output. 
\end{itemize}
\end{itemize}

\noindent\textbf{Security Level:} As noted in the previous work~\cite{liu2023vertical}, existing cryptographic defense mechanisms result in different levels of privacy during training, depending on what information is exposed. In alignment with the privacy level classification outlined in ~\cite{liu2023vertical}, we define the security levels of cryptogprahic defense mechanisms in ascending order of privacy protection as follows:
\begin{itemize}
    \item \textbf{S-1}: This level aims at protecting intermediate data exchanged between parties during training using cryptographic protocols, while partial training information could be still revealed. In this case, even if no intermediate data is disclosed, a party may still have plaintext access to its local model parameters and their updates during training, such as aggregated gradients with respect to the local model.
    \item \textbf{S-2}: Nothing is revealed to a party except their resulting trained local models. It is typically achieved through a trusted third party that provides the crypto service, or TEEs, or by employing HE and MPC techniques.
    \item \textbf{S-3}: Nothing is disclosed to a party, including the resulting trained local model. Similarly to S-2, S-3 is commonly achieved through a trusted third party, or TEEs or by employing HE and MPC-based techniques. In this case, model inference is either conducted over the encrypted data and encrypted local model that the owner party cannot decrypt, or by excluding the party and hosting its local model within a trusted domain.
\end{itemize}

\begin{table}[]
\caption{Security Level v.s. Knowledge. \cmark indicates what type of knowledge is protected at each security level.} \label{tab:secvsknowledge}
\small
\begin{tabular}{|c|c|c|c|c|c|c|c|}
\hline
    & K-C & K-A & K-G          & K-I          & K-S          & K-M          & K-H \\ \hline
S-1 &     &     &              & $\checkmark$ &              &              &     \\ \hline
S-2 &     &     & $\checkmark$ & $\checkmark$ &              &              &     \\ \hline
S-3 &     &     & $\checkmark$ & $\checkmark$ &              & $\checkmark$ &     \\ \hline
\end{tabular}
\end{table}

Table \ref{tab:secvsknowledge} presents the types of knowledge protected at each security level. Notably, S-1 level defenses, lacking protection against K-G type knowledge, may be vulnerable to related attacks. For instance, gradient matching attacks, as discussed in~\cite{zou2022defending}, can effectively infer private labels in VFL system with homomorphic encryption~\cite{yang2019federated}. Similarly, S-2 level defenses, which lack protection against K-M type knowledge, and S-3 level defenses, without protection against K-S type knowledge, may also have vulnerabilities. An example of such a vulnerability is the Model Completion attack~\cite{fu2022label}. This attack, assuming only the use of auxiliary samples and access to the local model for a label inference attack, can still be effective even in VFL systems that employ S-2 level defenses. Additionally, Model Inversion attacks such as \cite{gaopcat} can reconstruct input features from prediction scores, even in VFL systems protected by S-3 defenses, provided there are no restrictions on access to the prediction API.

\vspace{5pt}
\subsubsection{Non-cryptographic Defense Mechansims}~

\noindent\textbf{Differential Privacy (DP)} \quad
DP~\cite{dwork2006differential} is a mathematical framework for quantifying the privacy guarantees provided by an algorithm. In essence, it ensures that the removal or addition of a single database item does not significantly affect the outcome of any analysis, thus preserving the privacy of individual data. This is achieved by adding controlled random noise to the function outputs, which masks the contribution of individual data points. Local differential privacy~\cite{yang2020local} is a distributed variant of DP in which each individual applies the privacy mechanism to his data locally and sends the perturbed result to the central server.

\noindent\textbf{Data Obfuscation (DO)}\quad
Data obfuscation is a common strategy adopted by existing defense mechanisms to protect VFL data privacy, which typically involves randomly perturbed gradients, obfuscated model inputs and outputs, and fake labels. Obfuscated gradients are used to thwart gradient inversion attacks where an adversary attempts to reconstruct the input feature or label from the gradients. Some mechanisms use fake labels to defend label inference attacks.

\noindent\textbf{Adversarial Training (AT)}\quad  Adversarial training is another common strategy used by some defense mechanisms to mitigate privacy leakage. By incorporating adversarial objectives against inference attacks, it forces the model to learn to produce outputs that are less informative about the underlying sensitive data, thereby enhancing privacy. For example, previous work~\cite{sun2022label} proposed to limit the attacker’s ability to steal labels by reducing the correlation between forward embedding and target labels during training.

It is worth noting that these strategies are not mutually exclusive and can be combined with each other for defense.
In the following four sections, we first present privacy-preserving entity alignment in data pre-processing phase (Section \ref{sec:preprocess}), and then discuss attacks and defense strategies specific to each phase of a VFL lifecycle, including model training(Section \ref{sec:modeltraining}), model deployment(Section \ref{sec:modeldeployment}), and model inference(Section \ref{sec:modelinference}).

\section{VFL Phase: Data Prepossessing}
\label{sec:preprocess}
\begin{figure}
\includegraphics[width=0.4\textwidth]{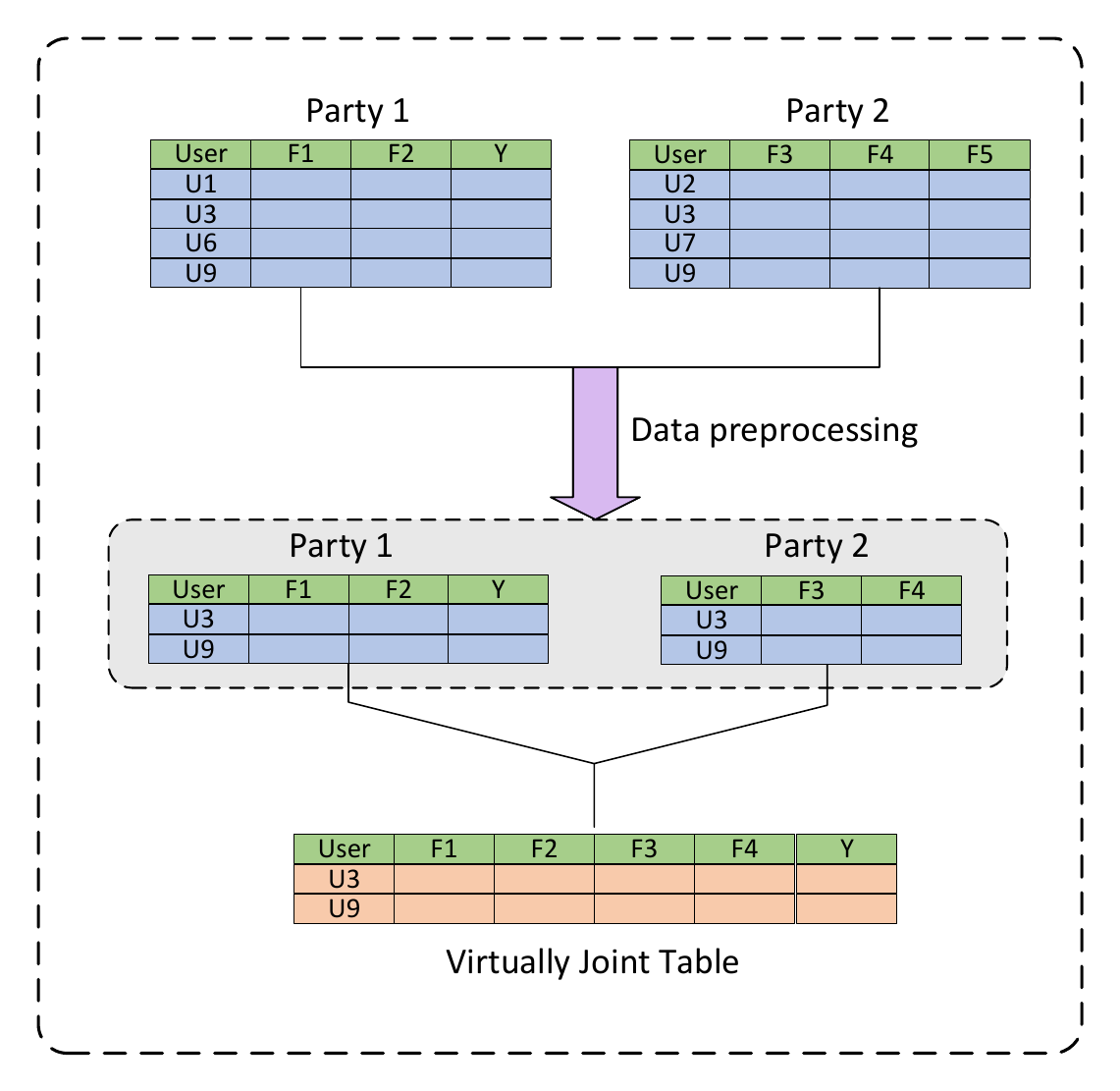}
\caption{Entity alignment in Data Processing phase.}
\label{dataprocessing}
\centering
\end{figure}

Data preprocessing is the initial stage of VFL that provides consistent data representation for the model training phase. This phase performs standard procedures like normalization, data cleaning, and feature selection. However, unique to VFL, is the critical process of entity alignment, which is not a concern in HFL. Figure \ref{dataprocessing} shows the entity alignment process on common user ids and the virtually joint features of two parties are used for VFL training. Without privacy-preserving measures, the process of aligning entities could reveal private data to other participants. Therefore, Private Set Intersection (PSI) protocols are used to construct secure entity alignment algorithms to find common sample IDs without revealing unaligned samples.

\noindent\textbf{Private Set Intersection (PSI)}\quad PSI is a cryptographic protocol used to identify the intersection of two sets of data in a secure and private manner, without actually revealing any additional information. It has been widely used for VFL entity alignment\cite{DBLP:conf/trustcom/LuD16, romanini2021pyvertical, cheng2020federated}. As shown in Fig. \ref{dataprocessing}, PSI uses the User ID as a unique identifier for the entity to obtain common sample IDs $\{U3, U9\}$.

Typical PSI protocols are designed for two-party scenarios. They are implemented based on various cryptographic techniques, such as Diffie-Hellman based PSI~\cite{DBLP:conf/sp/Meadows26, huberman1999enhancing}, oblivious transfer(OT) based PSI~\cite{DBLP:conf/uss/Pinkas4, DBLP:conf/ccs/KolesnikovKRT31, DBLP:journals/tissec/PinkasSZ6}, garbled circuit(GC) based PSI~\cite{huang2012private,DBLP:conf/eurocrypt/Pinkas0TY30,DBLP:conf/ccs/DCW1}, and algebraic PSI~\cite{freedman2004efficient,ghosh2019algebraic,kissner2005privacy,chongchitmate2022psi}.  
Because most of these protocols are designed with a semi-honest attack model, a line of works have been proposed to  strengthen PSI protocols against malicious attackers. Rosulek et al.~\cite{rosulek2021compact} improved the two-party PSI protocol with provable security against malicious parties in the ideal permutation and random oracle model. Pinkas et al.~\cite{pinkas2020psi} proposed a two-party PSI that uses Cuckoo hashing in 1-out-of-N OT extension based PSI to provide security against malicious participants. Dong et al.~\cite{dong2013private} presented a two-party PSI based on OT extension and oblivious Bloom intersectio, and further enhanced it against a malicious party by using symmetric key encryption of strings in the garbled Bloom filter. However, Rindal et al. \cite{rindal2017improved} demonstrated that the protocol~\cite{dong2013private} is not malicious-secure and fixed the bug with a lightweight cut-and-choose approach on OT messages. In \cite{rindal2017malicious}, they proposed a malicious-secure two-party PSI by carefully applying the dual execution paradigm of Mohassel et al. \cite{mohassel2006efficiency} to the OT-based PSI paradigm~\cite{DBLP:conf/uss/Pinkas4}. Ghosh et al.~\cite{ghosh2019algebraic} proposed a two-party PSI based on Oblivious Linear Function Evaluation (OLE) that encodes the elements in each set as roots of a polynomial. The authors showed the protocol is information-theoretically secure against a malicious adversary. Chongchitmate et al.~\cite{chongchitmate2022psi} constructed a two-party PSI protocol based on a ring version of oblivious linear function evaluation (ring-OLE) with security against malicious parties.

For multi-party scenarios, two-party PSI protocol can be directly used to obtain the intersection result, by computing the intersection with the next party sequentially until the last party. However, this method will leak intermediate intersection information. To address it, multi-party PSI protocols~\cite{freedman2004efficient,DBLP:conf/ccs/KolesnikovMPRT14,DBLP:conf/scn/InbarOP15} have been proposed. Freedman et al. \cite{freedman2004efficient} proposed a multi-party PSI protocol scheme based on HE. Kolesnikov et al. \cite{DBLP:conf/ccs/KolesnikovMPRT14} proposed a multi-party PSI protocol based on Oblivious Pseudo-Random Function (OPRF). Inbar et al. \cite{DBLP:conf/scn/InbarOP15} proposed two types of multi-party PSI protocols for semi-honest adversary and enhanced semi-honest adversary model respectively, based on OT and garbled Bloom filter~\cite{DBLP:conf/ccs/DCW1}. Lu et al. \cite{DBLP:conf/trustcom/LuD16} proposed a new multi-party PSI protocol in VFL, which can even handle the case that some parties drop out in the running of the protocol. Zhang et al. \cite{DBLP:conf/ccs/ZhangLLJL8} proposed PSI protocol for multiple parties based on the idea of a star network topology and Bloom filter. Ghosh et al.~\cite{ghosh2019algebraic} generalized their two-party PSI protocol to the fully malicious multiparty case, using only simple combinatorial techniques to ensure the security.

A potential risk with these PSI protocols is that they still reveal the common sample IDs to all parties. To address this problem, Liu et al. \cite{DBLP:journals/corr/abs10} introduce the notion of asymmetric VFL where a weak party holds a much smaller set of samples than a strong party and thus is more vulnerable to common ID disclosure. They proposed an asymmetric PSI protocol (APSI), which uses the Pohlig-Hellman encryption scheme and fuzzy set technique to protect the weak party from intersection information leakage in asymmetric VFL. Sun et al.~\cite{sun2021vertical} proposed a two-party Private Set Union (PSU) protocol that securely computes the union of sample IDs from two parties with alignment, which does not reveal the membership information of the intersection. For sample IDs that do not exist in the dataset of two parties, synthetic samples are generated for VFL training. Du et al.~\cite{DBLP:journals/corr/abs12} proposed to combine Diffie-Hellman PSI with differential privacy to limit the leakage of membership information from the output.

\section{VFL Phase: Model Training}\label{sec:modeltraining}
In this section, we examine the various types of privacy attacks that arise during the training phase of VFL, along with the defense mechanisms designed to mitigate them.
For model training stage, we focus on the attacks that target at the training dataset.
Based on the type of specific information targeted by privacy attacks, we will delve into label inference attacks, feature inference attacks and property inference attacks
in the following sections. For each type of attack, we review existing research organized by the
methodology they employed.


\subsection{Label Inference Attack}
This attack aims to deduce the actual labels of the training data held by individual participants, based on the gradients or intermediate results they transmit. Based on the information used by deduction methodologies, we categorize the existing work as gradient-based attacks and intermediate data-based attacks.

\subsubsection{Gradients Based Attacks }\label{sss:labelgradientattack}
The primary source of these privacy risks stems from the exchange of gradients, an essential part of the training process in federated learning. An adversary who can access these gradients could deduce the training data specific to an individual participant.

\indent\textit{1. Gradient Sign-based Inference}
Gradient Sign-based Inference, also called Direct Label Inference (DLI) attack~\cite{fu2022label}, is an attack that infers labels directly from the signs of gradients. It assumes the scenario of VFL without model splitting, where the passive parties send logits to the active party, and the active party computes loss with its label data.  The adversary, who is a passive party in VFL, can directly infer the label information held by the active party based on the sign of gradients of the final prediction layer received from the active party. Suppose the model uses cross-entropy loss,
the gradient of the loss w.r.t. the $i$-th logit from the adversary is
\begin{equation}
    g_{i}^{a d v}=\frac{\partial \operatorname{loss}(x, c)}{\partial y_{i}^{a d v}}
\end{equation}
where $c$ is the ground truth label. If $g_{i}^{a d v}<0$ if $i=c$ and $g_{i}^{a d v}>0$ if $i \neq c$.
Therefore, by checking the sign of the gradients received from the server, the attacker can infer the labels of its local training dataset. DLI is also applicable to other types of loss function such as cross-entropy loss, weighted cross-entropy loss, and negative log-likelihood loss.

\textit{2. Gradient Norm Scoring-Based Inference.}
Li et al.~\cite{li2021label} consider label inference within the scenario of two-party split learning binary classification. The proposed attack exploits the fact that the gradient norm of the positive instances is generally larger than that of the negative ones.

\textit{3. Gradient Direction Scoring-Based Inference.} With the same problem setup as above, Li et al.~\cite{li2021label} also proposed another attack, based on the cosine similarity of two gradient vectors from two examples. For a target example, all examples in the same class result in positive cosine similarity, whereas all opposite-class examples have negative cosine similarity. Knowing the label of an example in the same group, the passive party can determine the label of the target example. Because real-world datasets usually have the class-imbalance problem, the attacker can determine the label if he is aware of which class has more/less examples.


\indent\textit{4. Gradient Clustering-based Inference.}
Liu et al. \cite{liu2022clustering} proposed an inference attack by applying clustering to the gradients received from the active party during the backpropagation process during the training phase. Clustering uses cosine similarity as the distance metric and $K$-means algorithm. The attack maps $K$ clusters to $K$ labels with the auxiliary data.

\indent\textit{5. Gradient Matching-based Inference.}
In general, attacks based on gradient matching infer labels by looking for labels that can produce the gradients closest to those received from the active party or coordinator server during training.
The Unsplit attack proposed by Erdogan et al.~\cite{erdougan2022unsplit} assumes that the passive party knows the model architecture of the active party. By computing the gradient values resulting from backpropagation with a clone model for each possible label, the attacker determines the label as the one that produces the closest gradients to the gradients received from the active party.
The ExPLoit attack~\cite{kariyappa2023exploit} infers the labels by solving a supervised learning problem. It trains a surrogate model along with surrogate label variables, with a gradient matching objective that minimizes the $l_2$ distance between the gradients and the original gradients received from the active party. The training objective also includes regularization terms for label inference that minimize the entropy of the surrogate label and the distance between the distribution of the surrogate labels and the prior expected label distribution of the training data.
Qiu et al. \cite{qiu2023exact} proposed a data reconstruction attack named EXACT for split learning in which the gradients of the activations are transmitted from the client to the server. The attack assumes access to the client-side model parameters during training and aims to recover private data at the client, which can be either private labels or private features. This attack assumes that the client's private labels/features are categorical or can be divided into a finite number of categories. It constructs a list of all possible combinations of features and labels, each referred to as a configuration. The attack computes the gradients under each configuration and chooses the configuration that minimizes the $l_2$ distance to the true gradient returned by the client.

Assuming that the VFL training protocol is protected by homomorphic encryption and participants only receive batched-averaged local gradients in plaintext from trusted coordinator server, Liu et al. \cite{liu2021batch,zou2022defending} proposed an attack that allows label inference for a batch of examples based on averaged gradients. The adversarial passive party deduces the labels of samples in a batch by minimizing the discrepancy between the received average gradients and the simulated gradients with respect to the variables of the label and intermediate result corresponding to every sample in the batch.

\indent\textit{6. Gradient Linear Inversion Inference.} 
Hu et al.~\cite{hu2022vertical} analyzed label privacy for the active party in vertical logistic regression. During the training time, the passive party $B$ receives the gradients w.r.t its local model, which is the linear combination on its private features. Assuming that the batch size $s$ is not greater than $B$'s feature dimension $d_B$, Party $B$ can determine the label based on the sign of the combination coefficients when the initialization of the model weight is small. The authors also derived the number of training iterations during which the criterion of using the coefficient sign can be valid for label inference. Furthermore, the authors developed an active attack under the HE training protocol to make the label inference more practical. The idea is to maliciously add dummy features for $B$ to make $s \le d_B$ by generating and compressing the auxiliary ciphertext sent from $B$ to $A$ for the training with HE.

Tan et al. \cite{tan2022residue} proposed a residual variable-based label inference attack method for binary classification of two-party vertical logistic regression (LR). The attacking party uses the decrypted gradients w.r.t its local model and data features to formulate a set of linear equations, aiming to solve for the unknown encrypted residue vector. In LR, the residue value is calculated as the difference between the ground truth label and the value of the sigmoid function applied to the input features. Using this information, the attacker can infer that the actual true label is 1 if the residue $r>0$, and 0 otherwise.

\subsubsection{Intermediate Result Based Attacks}
\label{sss:labelmedattack}
In contrast to gradient-based attacks, intermediate data-based attacks do not directly use gradients but intermediate results, such as embedding results during the forwarding process, to recover the label information.

\textit{1. Spectrum-based Attack.}
Sun et al.~\cite{sun2022label} proposed an attack that infers private labels from the shared intermediate embedding for binary classification. It leverages spectral attack~\cite{tran2018spectral} which is a single value decomposition (SVD) based outlier detection method. The spectral attack is used to differentiate the cut-layer embedding distribution between the positive samples and the negative samples. The attacker uses the auxiliary information of the population-level
information (i.e. ratio of the positive instances) regarding the positive and negative class’s cut-layer embedding distributions, in order to determine which cluster belongs to which label.
The authors show that this attack can still achieve label inference despite the implementation of defense strategies such as Marvell \cite{ghazi2021deep} and label DP \cite{ghazi2021deep}, which we will introduce in Section \ref{ssec:defense}.

\textit{2. Instance Space based Attack.}
Takahashi et al. \cite{takahashi2023eliminating} proposed a label inference attack named ID2Graph against Tree VFL, by exploiting the instance spaces of tree nodes, i.e., the set of sample IDs assigned to each tree node. During training, the active party shares the instance space of a node with passive parties for further splitting. ID2Graph assumes that the samples within the same instance space have similar class labels. Based on that, ID2Graph deduces private training labels by clustering the sample IDs into groups of the same number of class categories. A passive party attacker first constructs a graph representation for sample IDs exposed to him, where each vertex is a sample, and an edge means that two vertices belong to the same leaf. It runs community detection on the graph and uses community allocation along with its local data features for sample clustering. The attacker can determine the training labels if it has label distribution information or any labeled sample in a cluster.

\subsection{Feature Inference Attack}
This type of attack aims to deduce the private training data or attributes of the training examples held by passive participants.

\subsubsection{Gradient Based Attacks}
\label{sss:gradfeatureattacks}
The gradients based attacks aim to recover the features of training data from the gradients exchanged in VFL.

\indent\textit{1. Gradient Matching-Based Inference.} 
The gradient matching-based attacks aim to recover the training data by optimizing the dummy gradients computed on the dummy data as close as to the received original gradients computed on the real data, which also makes the dummy data close to the real training data. It is first proposed in DLG attack \cite{zhu2019deep} for horizontal federated learning and distributed training. Suppose a selected batch $B=\{(x_n, y_n)\}$. the recovered batched data $B'=\{(x'_n, y'_n)\}$ can be obtained by optimizing the objective function
\begin{equation}
\label{eq:gradientopt}
    B' = \arg\min\limits_{B'} \norm{\frac{1}{|B|}\sum\limits_{B} \nabla_\Theta \mathcal{L}(\Theta,x_n,y_n)-\frac{1}{|B|}\sum\limits_{B'} \nabla_\Theta \mathcal{L}(\Theta,x'_n,y'_n)}^2
\end{equation}

The EXACT attack by Qiu et al. \cite{qiu2023exact}, which we have discussed in Section \ref{sss:labelgradientattack}, assumes access to the client-side model parameters and uses gradient matching to recover the private features of the client in split learning.
Jin et al. \cite{NEURIPS2021_08040837} proposed an attack CAFE that leverages data index alignment and internal representation alignment in VFL to address the batch size problem, because as the batch size is sufficiently large, it will be more challenging to find the ``right'' solution $B'$. 
CAFE addresses this problem with three steps. First, by iterative optimization, it separates the gradients of loss w.r.t each single input to the first (fully connected) FC layer from the aggregated gradients. Next, it recovers the inputs to the first FC layer by minimizing the difference between the gradients derived from the inputs and the gradients obtained from the first step. Lastly, it initializes pseudo-data and pseudo-labels randomly and recovers real data by matching gradients along with the regulararizer that matches internal representation. CAFE demonstrates the capability of recovering large-batch size data in general VFL protocols. DLG has a batch size limitation of $8$, while CAFE can recover data up to the maximum batch size of $100$. 

The data reconstruction algorithms based on Equation~\ref{eq:gradientopt} rely on the Euclidean distance and optimization via L-BFGS, which could not be optimal since the distance only captures the magnitude not direction. Therefore, Geiping et al. \cite{geiping2020inverting} proposed using cosine similarity between gradients as a loss function to reconstruct the data.

\indent\textit{2. Gradient Linear Inversion Inference.} 
In the work on Gradient Linear Inversion for Label Inference~\cite{hu2022vertical} we discussed before, Hu et al. also analyzed the feature privacy for the passive party $B$ in vertical logistic regression. During training time, the active party $A$ receives the gradients w.r.t its local model, which is the linear combination on its private features. When the batch size $s$ is smaller than $A$'s feature dimension $d_A$, Party $A$ can determine the coefficients of the combination by solving a linear equation system. These coefficients can then be used to derive the ratio among the data items of the party $B$ when $B$ only has a single feature, that is, $d_B=1$. The authors also provided the hardness result for feature inference when $d_B\ge 2$. The proposed active attack can also be applied to ensure $s \le d_A$ thereby making the feature inference practical in HE training.

\subsubsection{Intermediate Result Inversion Attacks}
\label{ssec:resdataintraining}
This type of attack uses intermediate results during training to infer private training data using the intermediate results (e.g., forwarded embeddings, or sometimes called smashed data) received from clients (that is, the passive party). It can be achieved through two approaches: 1. Equation Solving attacks that use intermediate data to construct and solve a system of equations to recover private input features; 2. Machine Learning-based attacks that use intermediate data to train a model to private infer input features. 

\textit{1. Equation Solving Attacks.} 
Ye et al. \cite{ye2022feature} introduced an inference attack in which the active party aims to reconstruct binary features of the passive party, based on intermediate results received from the passive party. It recovers the binary input features $\mathbf{x}\in\{0,1\}^n$ by solving the linear equations $\mathbf{Aw}=\mathbf{x}$ where $A$ is derived from the intermediate results. Furthermore, the authors proposed solving the linear regression
$$\argmin_{x} \min_w||\mathbf{Aw}-\mathbf{x}||^2_2$$ to avoid the unreliability problems of solving linear equations. This attack could be performed in any training round.

Weng et al. \cite{weng2020privacy} proposed a reverse multiplication attack against two-party secure logistic regression VFL in which both participants train a global model and require a third-party coordinator to aggregate encrypted partial gradients. This attack has the strong assumption that the adversary has the private key of the coordinator. It uses the intermediate output of another party and decrypted gradients to construct a system of linear equations and compute the unknown features by solving them. 
Furthermore, they also proposed a reverse sum attack against SecureBoost~\cite{kewei2021secureboot} in which the two participants independently learn a joint model without any third party. It is an active attack that aims to determine the order of the samples for the features of the passive party. The adversary encodes a random magic number into the first and second gradients and uses the pattern to reverse engineer all the addition terms from the gradient sum received from the passive party, which reveal the partial orders of features.

\textit{2. Surrogate Model Learning-based attacks.}
The Unsplit attack~\cite{erdougan2022unsplit}, which we discussed in the label inference attack, also proposed a model inversion attack to recover the input data of the client. Given the knowledge of the model architecture of the passive party, the attacker applies "coordinated gradient descent" to iteratively update a surrogate clone model and the estimated input values by minimizing the distance between the true client's output and the clone model's output. The final result provides the estimation of the input data.

\textit{3. Generative model based attacks.}
Pasquini et al. \cite{pasquini2021unleashing} proposed a Feature-space Hijacking Attack (FSHA). Compared with Unsplit attack~\cite{erdougan2022unsplit}, it is an active attack which means that the adversary actively controls the learning process. In this attack, the malicious active party has access to an auxiliary dataset $X_{pub}$ that is from the same domain of the training data of the passive parties. During the VFL training process, it trains an auto-encoder function consisting of an encoder network $f'$ and a decoder network $f'^{-1}$ with $X_{pub}$, and simultaneously trains a discriminator $D$ to distinguish between the feature
space induced from $f'$ and the one induced from the passive party’s network $f$. The key of FSHA is to send to the passive party forged gradients that is to maximize the probability of $f$'s output being misclassified by $D$. In essence, the passive party is forced to map its input to a feature space that is indistinguishable from that of $f'$. $f^{-1}$ is trained to reach a state that allows the attacker to recover private training instances from smashed data.

\subsection{Model Extraction Attack.}
\label{ssec:meintraining}
During model training, an adversary has the potential not only to infer the training data, but also to extract models from the exchanged information. The general approach to model extraction is to train a surrogate model that closely replicates the target model. The active party can perform model extraction against the passive party, and the reverse is also possible.  We discuss both scenarios in the following.

\subsubsection{Active Party Steals from Passive Party.}
\label{ssec:acpameintraining}
The Unsplit work~\cite{erdougan2022unsplit} proposed a model extraction attack in which the active party aims to steal the passive party model. It assumes knowledge of the passive party's model architecture. It initializes a surrogate model that has the same architecture and trains the model to minimize the distance between its output and the intermediate output received from the passive party during the split training time.

Gaopcat et al. \cite{gaopcat} proposed a pseudo-client attack (PCAT) that trains a surrogate model to the passive party's model. PCAT assumes that the server does not know anything about the client model structure but has access to an auxiliary dataset from the same domain. In each training iteration, the surrogate model is updated by gradient descent to minimize the KL divergence between two soft labels produced with the client model and the surrogate model, respectively.

\subsubsection{Passive Party Steals from Active Party.}
Li et al. \cite{li2023model} proposed several attack approaches to extract the model of the active party, where the attacker holds a gray-box assumption, that is, it knows the architecture and loss function while the model parameters are unknown. The Craft-ME attack assumes that the passive party can create a number of low-loss instances for each label. This is achieved by iteratively updating an input from a random initialization with gradient updates received from the active party. Then, it uses these small-loss instances to train a surrogate model of the active party's model. 
The GAN-ME attack instead uses a GAN-based approach to generate small-loss instances and trains the generator during VFL training by generating fake data and labels and obtaining gradient updates from the active party. The GM-ME (Gradient Matching) attack uses auxiliary labeled data and trains the surrogate model with the objective to match the surrogate model's gradients and the gradients received from the active party. 
The Train-ME attack assumes that the attacker has access to a subset of training data and uses it to train an accurate surrogate model.
The Soft-train-ME (Gradient-based soft label training) attack also assumes the availability of a subset of training data. The gradients on ground truth label and incorrect labels are used to compute the soft labels which are then used to train the surrogate model. Note that these attacks operate under the assumption that 1. the attacker can query the active party to obtain the gradient of any labeled input, and 2. the active party blocks the prediction queries so that neither logits nor prediction labels are accessible by passive parties during training, which may not be practical in common VFL settings.


Table \ref{tab:attacksintraining} summarizes all attacks that we reviewed during the training phase. They are all insider attacks and have access to local model parameters. We use explicit `K-M' in the `Knowledge' column to indicate that an attack assumes that the adversary has access to the complete model or to the models belonging to other parties.
The `Attack Direction' column indicates the attack direction between passive party denoted by `P' and active party denoted by 'A'. The party on the left side of '->' represents the attacker, while the right side is the victim. In cases where there is no distinction between an active and passive party, this is indicated by `\textbackslash'.
The 'Attacker' column categorizes the behaviors of attackers as either `P' for passive or `A' for active attacker followed by the attacker's capability. 
The `\# parties' column specifies the number of parties involved in the VFL setting where the attack is developed.
The `HE protocol' column specifies whether the attack is explicitly designed for use under the VFL training protocol that employs HE.

\begin{table}[]
\caption{Summary of existing attacks in model training phase of VFL.}\label{tab:attacksintraining}
\scriptsize
\setlength{\tabcolsep}{1pt}
\begin{tabular}{|c|c|c|c|c|c|c|c|c|c|c|}
\hline
Attack                & Existing Attacks                     & \begin{tabular}[c]{@{}c@{}}Attack \\ Strategy\end{tabular} & Method                                                                                        & Knowledge     & \begin{tabular}[c]{@{}c@{}}Attack \\ Direction\end{tabular} & \begin{tabular}[c]{@{}c@{}}Attack \\ Type\end{tabular} & \begin{tabular}[c]{@{}c@{}}Attacker \\ (Capability)\end{tabular} &
\# parties &
\begin{tabular}[c]{@{}c@{}}Model\\ (\# classes)\end{tabular} & \begin{tabular}[c]{@{}c@{}}HE \\ proto\end{tabular} \\ \hline
\multirow{13}{*}{LIA} & DLI\cite{fu2022label}                & G-S                                                        & Gradient Sign                                                                                 & K-G           & P->A                                                        & B-A                                                      & P                                               &        $\ge 2$            & NN (>=2)                                                     & N                                                      \\ \cline{2-11} 
                      & \cite{li2021label}                   & G-S                                                        & Gradient Norm Scoring                                                                         & K-G, K-C      & P->A                                                        & B-A                                                      & P                          &                   2                      & NN (2)                                                       & N                                                      \\ \cline{2-11} 
                      & \cite{li2021label}                   & G-S                                                        & Gradient Direction Scoring                                                                    & K-G, K-C      & P->A                                                        & B-A                                                      & P                          &                    2                     & NN (2)                                                       & N                                                      \\ \cline{2-11} 
                      & \cite{liu2022clustering}             & G-S                                                        & Gradient Clustering                                                                           & K-G           & P->A                                                        & B-A                                                      & P                           &                   2                     & NN ($\ge 2$)                                                     & N                                                      \\ \cline{2-11} 
                      & Unsplit\cite{erdougan2022unsplit}    & G-S                                                        & Gradient Matching                                                                             & K-G, K-H      & P->A                                                        & G-A                                                      & P                            &                  2                     & NN ($\ge 2$)                                                     & N                                                      \\ \cline{2-11} 
                      & ExPLoit\cite{kariyappa2023exploit}   & G-S                                                        & Gradient Matching                                                                             & K-G, K-C      & P->A                                                        & B-A                                                      & P                            &                   2                    & NN ($\ge 2$)                                                     & N                                                      \\ \cline{2-11} 
                      & EXACT\cite{qiu2023exact}             & G-S                                                        & Gradient Matching                                                                             & K-G, K-M, K-C & \textbackslash                                                           & W-A                                                      & P              &                   2                                  & NN ($\ge 2$)                                                     & N                                                      \\ \cline{2-11} 
                      & \cite{liu2021batch,zou2022defending} & G-S                                                        & Gradient Matching                                                                             & K-G           & P->A                                                        & B-A                                                      & P                             &                    $\ge 2$                  & NN ($\ge 2$)                                                     & Y                                                      \\ \cline{2-11} 
                      & \cite{hu2022vertical}                & G-S                                                        & Gradient Linear Inversion                                                                     & K-G           & P->A                                                        & B-A                                                      & P                              &                    $\ge 2$                 & LR ($\ge 2$)                                                     & Y                                                      \\ \cline{2-11} 
                      & \cite{hu2022vertical}                & G-S                                                        & \begin{tabular}[c]{@{}c@{}}Gradient Linear Inversion \\ for HE training\end{tabular}          & K-G           & P->A                                                        & B-A                                                      & A(IO-M)                         &                     2               & LR ($\ge 2$)                                                     & Y                                                      \\ \cline{2-11} 
                      & Residual\cite{tan2022residue}        & G-S                                                        & Gradient Linear Inversion                                                                     & K-G           & P->A                                                        & B-A                                                      & P                               &                      2              & LR (2)                                                       & Y                                                      \\ \cline{2-11} 
                      & \cite{sun2022label}                  & I-S                                                        & Spectrum Attack                                                                               & K-I, K-C      & P->A                                                        & B-A                                                      & P                                &                    2               & NN (2)                                                       & N                                                      \\ \cline{2-11} 
                      & \cite{takahashi2023eliminating}      & I-S                                                        & Instance Space                                                                                & K-I, K-C      & P->A                                                        & B-A                                                      & P                                &               $\ge 2$                    & Tree ($\ge 2$)                                                   & N                                                      \\ \hline
\multirow{9}{*}{FIA}  & EXACT\cite{qiu2023exact}             & G-S                                                        & Gradient Matching                                                                             & K-G, K-M, K-C & \textbackslash                                                           & W-A                                                      & P                                        &               2            & NN ($\ge 2$)                                                     & N                                                      \\ \cline{2-11} 
                      & CAFE\cite{NEURIPS2021_08040837}      & G-S                                                        & Gradient Matching                                                                             & K-G, K-M      & A->P                                                        & W-A                                                      & P                                  &               $\ge 2$                  & NN ($\ge 2$)                                                     & N                                                      \\ \cline{2-11} 
                      & \cite{hu2022vertical}                & G-S                                                        & Gradient Linear Inversion                                                                     & K-G           & A->P                                                        & B-A                                                      & P                                   &               $\ge 2$                 & LR ($\ge 2$)                                                     & Y                                                      \\ \cline{2-11} 
                      & \cite{hu2022vertical}                & G-S                                                        & \begin{tabular}[c]{@{}c@{}}Gradient Linear Inversion \\ for HE training\end{tabular}          & K-G           & A->P                                                        & B-A                                                      & A(IO-M)                              &                  2             & LR ($\ge 2$)                                                     & Y                                                      \\ \cline{2-11} 
                      & \cite{ye2022feature}                 & I-S                                                        & Equation Solving                                                                              & K-I           & A->P                                                        & B-A                                                      & P                                    &            2                   & NN ($\ge 2$)                                                     & N                                                      \\ \cline{2-11} 
                      & \cite{weng2020privacy}               & I-S                                                        & \begin{tabular}[c]{@{}c@{}}Equation Solving \\ (reverse multiplication)\end{tabular}          & K-I, K-G      & A->P                                                        & B-A                                                      & P                                     &            2                  & LR ($\ge 2$)                                                     & Y                                                      \\ \cline{2-11} 
                      & \cite{weng2020privacy}               & I-S                                                        & \begin{tabular}[c]{@{}c@{}}Equation Solving\\ (reverse sum)\end{tabular}                      & K-I, K-G      & A->P                                                        & B-A                                                      & A(G-M)                                 &            2                 & LR ($\ge 2$)                                                     & Y                                                      \\ \cline{2-11} 
                      & Unsplit\cite{erdougan2022unsplit}    & S-S                                                        & Surrogate Model Learning                                                                      & K-I, K-H      & A->P                                                        & G-A                                                      & P                                      &            2                 & NN ($\ge 2$)                                                     & N                                                      \\ \cline{2-11} 
                      & FSHA\cite{pasquini2021unleashing}    & I-S                                                        & GAN+Adversarial Training                                                                          & K-I, K-A      & A->P                                                        & B-A                                                      & A(L-M)                                &            2                  & NN ($\ge 2$)                                                     & N                                                      \\ \hline
\multirow{6}{*}{MEA}  & Unsplit\cite{erdougan2022unsplit}    & S-S                                                        & \begin{tabular}[c]{@{}c@{}}intermediate result \\ divergence loss\end{tabular}                & K-I, K-H      & A->P                                                        & G-A                                                      & P                                           &         2               & NN ($\ge 2$)                                                     & N                                                      \\ \cline{2-11} 
                      & PCAT\cite{gaopcat}                   & S-S                                                        & \begin{tabular}[c]{@{}c@{}}soft-label prediction\\ divergence loss\end{tabular}               & K-S, K-A      & A->P                                                        & B-A                                                      & P                          &               $\ge 2$                          & NN ($\ge 2$)                                                     & N                                                      \\ \cline{2-11} 
                      & Craft-ME\cite{li2023model}           & S-S                                                        & \begin{tabular}[c]{@{}c@{}}small-loss training examples\\ crafted from gradients\end{tabular} & K-G           & P->A                                                        & G-A                                                      & P                           &                $\ge 2$                        & NN ($\ge 2$)                                                     & N                                                      \\ \cline{2-11} 
                      & GAN-ME\cite{li2023model}             & S-S                                                        & \begin{tabular}[c]{@{}c@{}}small-loss training examples\\ generated by GAN\end{tabular}       & K-G           & P->A                                                        & G-A                                                      & P                            &               $\ge 2$                        & NN ($\ge 2$)                                                     & N                                                      \\ \cline{2-11} 
                      & GM-ME\cite{li2023model}              & S-S                                                        & Gradient Matching                                                                             & K-G           & P->A                                                        & G-A                                                      & P                             &              $\ge 2$                        & NN($\ge 2$)                                                      & N                                                      \\ \cline{2-11} 
                      & Soft-train-ME\cite{li2023model}      & S-S                                                        & \begin{tabular}[c]{@{}c@{}}Gradient-based \\ soft-label training\end{tabular}                 & K-A, K-G      & P->A                                                        & G-A                                                      & P                              &              $\ge 2$                       & NN ($\ge 2$)                                                     & N                                                      \\ \hline
\end{tabular}
\end{table}

\subsection{Defenses}\label{ssec:defense}


In this section, we discuss defense methods against privacy attacks during the model training stage, according to the category of their methodologies.

\subsubsection{Cryptographic Defense}~

\noindent\textbf{Homomorphic encryption (HE) based defense}\quad  Applying HE in VFL training protocol can effectively defeat most attacks we have discussed that exploit gradient information and intermediate results received from the victim party, because these information are not accessible to the attacker and the attacker can only obtain decrypted gradients with respect to its own local model parameters during training.
Many HE based defense schemes have been proposed to secure VFL, targeting linear and logistic regression, decision trees, and neural network models. Most of them require a trusted coordinator for key distribution and decryption. We discuss them by their targeted models. 

\begin{figure}[H]
\includegraphics[width=0.3\textwidth]{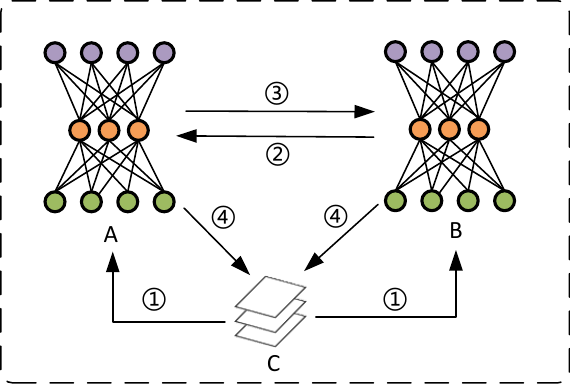}
\caption{A is an active party, B is a passive party, and C is a coordinator. step\textcircled{1}:sending public keys;
step\textcircled{2}:sending intermediate results;  step\textcircled{3}:sending loss;step\textcircled{4}:sending encrypted gradients;step\textcircled{5}:sending decrypted gradients}
\label{VFLHE}
\centering
\end{figure}

\indent\textit{VFL linear/logistic regression:}\quad
A typical HE-based framework for secure VFL linear/logistic regression is shown in Figure \ref{VFLHE}. Its main idea is to encrypt the intermediate results and compute the loss and gradients on the ciphertext. The coordinator decrypts the gradients for each participant.
The additive HE-based linear regression proposed by Yang et al.~\cite{yang2019federated} for VFL is to learn the model $\mathbf{Y}=\Theta \mathbf{X}$. Suppose that $A$ holds samples $\{X_A, Y\}$ and $B$ holds $X_B$ where the model paramter $\Theta = (\Theta_A, \Theta_B)$ each corresponding to the feature space of $X_A$ and $X_B$ respectively. The coordinator $C$ distributes the
public key used by HE to parties $A$ and $B$.
Party $B$ computes $[[\Theta_BX_B]]$ and sends it to $A$. Party $A$ computes $[[U]]=[[Y-\Theta_AX_A-\Theta_BX_B]]$ and sends it to $B$. $A$ also computes the gradients $[[g_A]]==[[U]]X_A$, adds a random mask $[[R_A]]$ and sends the result $[[g_A+R_A]]$ to the coordinator $C$. Party $B$ does the same with adding a random mask to gradients and sending the result $[[U]]]X_B+[[R_B]]$ to $C$. The coordinator $C$ decrypts the results and sends $g_A+R_A$ to $A$ and $g_B+R_B$ to $A$ and $B$ respectively. $R_A$ and $R_B$ are used to hide true gradients from $C$. Using an architecture of non-colluding two servers holding outsourced data, Giacomelli et al. \cite{giacomelli2018privacy} proposed a privacy-preserving approach that trains a ridge linear regression model using only linearly-homomorphic encryption (LHE). It avoids using Yao's GC in the VFL LR protocol proposed by Gascon et al \cite{gascon2016secure}, by using the linear homomorphic property of LHE to solve linear equations for reducing the communication overhead.  

For VFL logistic regression, the HE-based scheme named HardyLR proposed by Hardy et al.~\cite{hardy2017private} shares a similar framework but with the Taylor approximation to apply HE to logistic regression and stochastic gradient descent. Yang et al proposed BaiduLR \cite{yang2019parallel} that removes the coordinator using party $A$ to distribute the public key to $B$ and decrypt $g_B+R_B$ for $B$. A key difference is that $B$ sends intermediate results $\Theta_BX_B$ in plaintext to $A$, which means that $A$ can accumulate sufficient equations through training iterations and solve the system of equations to infer $B$' private data. SecureLR \cite{he2021secure} addressed this problem by combining piecewise function and blind millionare algorithm to securely compute $U$ without the exchange of plaintext information. To resolve the slow convergence of first-order stochastic gradient descent and reduce the communication cost of VFL logistic regression, Yang et al.~\cite{yang2019quasi} developed a quasi-newton method for two-party VFL training under HE.
For these HE-based schemes, the previous work by Liu et al. \cite{liu2021batch,zou2022defending} demonstrates that even with HE, the attack may only use decrypted local gradients for label inference attacks. 

\indent\textit{VFL Decision Tree:}\quad
A number of HE-based privacy-preserving solutions have been proposed for training tree models in VFL, targeting gradient-boosting decision tree algorithms~\cite{natekin2013gradient} such as LightGBM~\cite{ke2017lightgbm} and XGBoost (XGB) \cite{chen2015xgboost}.

Hou et al. \cite{hou2021verifiable} proposed a verifiable privacy-preserving scheme VPRF for random forest (RF) in VFL which uses a coordinator server and involves a trusted authority to distribute keys for HE. To protect the labels and Gini coefficients, VPRF uses HE to encrypt them when exchanged between participants and the server. In addition, it allows participants to verify the correctness of the computation results received from the server. Liu et al. \cite{liu2021revfrf} proposed a revocable scheme named RevFRF for RF in VFL that can further protect the data privacy of a participant when it is revoked. During tree construction, the active party, referred to as a central server with ground truth in the paper, collects binary split results produced with candidate splits picked by participants, selects the best splits, and asks the contributing participants to send encrypted split information. Xu et al. \cite{XU2023VFCART} proposed a secure VFL framework for classification and regression tree called VF-CART where the active party generates keys of HE and sends the public key to a coordinator server. The server receives feature histograms received from all participants, interacts with the active party under HE encryption, and stores the split information of the tree with encrypted labels at leaf nodes. Although RevFRF and VF-CART mitigate the disclosure of private data among participants, they still partially expose information about features to the server due to the transmission of plaintext histograms and binary split results.

Feng et al. \cite{feng2019securegbm} proposed SecureGBM, a secure multi-party Gradient Boosting framework based on LightGBM. It uses semi-homomorphic encryption offered by Microsfot SEAL to implement addition and multiplication operators used in the LightGBM algorithm while safeguarding the binary comparison process for decision making within a trusted, isolated domain.
SecureBoost \cite{cheng2021secureboost} is a privacy-preserving tree-boosting that adapts XGBoost in VFL with HE. The active party acts as a trusted party and uses HE to encrypt the gradient statistics sent from the active party to the passive parties, to prevent label disclosure to passive parties. Each passive party computes gradient statistics under HE for all possible splits locally and sends them to the active party. The active party deciphers all encrypted gradient statistics from all parties and computes the global optimal split. In addition, the SecureBoost work analyzed potential label leakage caused by the weight of the leaves of the first tree in the model, and proposed Reduced Leakage SecureBoost (RL-SecureBoost) to reduce the risk through only learning and storing the first tree locally at the active party. Based on SecureBoost, Chen et al. introduced a framework called SecureBoost+ \cite{chen2021secureboost+}, which aims to reduce ciphertext-related computation and communication costs. It applies gradient packing and cipher compression to reduce the ciphertext size and incorporates a hybrid tree and layered tree training mode in the training phase to skip partial computation and communication operations. Zhang et al. \cite{zhang2023embedded} proposed a VFL feature selection algorithm based on particle swarm optimization under the SecureBoost framework. Fu et al. \cite{fu2021vf2boost} analyzed SecureBoost's performance bottlenecks caused by frequent mutual-waiting between parties and the cost of cryptography-related operations. They proposed a concurrent training system Vf2boost to reduce idle waiting time and customized cryptography operations to reduce the number of scaling operations.
Chamani et al. \cite{chamani2020mitigating} pointed out that in SecureBoost the the active party can still learn all splits of all features for every party and estimate the partial ordering across features of the passive parties. They proposed to adopt Trusted Execution Environment (TEE) at the active party to decrypt all splits and compute the optimal split.

\indent\textit{VFL Neural Network:}\quad  In the work of Zhang et al. \cite{zhang2020additively}, a privacy preservation scheme ACML was introduced for two-party VFL. It uses PHE to encrypt intermediate results during forward propagation and applies random noise to protect gradients during backward propagation. Liu et al. \cite{liu2020secure} proposed a two-party federated transfer learning approach HFTL where each party computes its local gradients with HE using the other party's public key and adds random noise before sending these gradients to the other party for description. However, it does not protect the label information because the prediction of private samples of one party is performed by the other party.
Kang et al. \cite{kang2022privacy} proposed PrADA that is similar to ACML but extended for federated adversarial domain adaptation. A third party acts as a bridge for transferring knowledge from the source to the target party with its complementary sample features and uses PHE to encrypt intermediate results to be exchanged. Other parties do the computation under HE, perturb the result with private random noise, and send it to the third party for decryption. Cheung et al. \cite{cheung2021fedsgc} considered VFL for graph neural networks where the graph structure and node features are stored separately in two parties. It uses the same strategy that combines HE and random perturbation to prevent privacy leakage from information exchanged between two parties. Yang et al. \cite{yang2023securesl} proposed SecureSL, a privacy-preserving VFL framework for Web 3.0 that uses multi-key HE (MKHE) to ensure the confidentiality of the data, the labels and the trained model. Khan et al. \cite{khan2023split} proposed a protocol based on U-shaped split learning for the 1D CNN model, which applies HE to the activation mapping on the client side before sending it to the server to protect user privacy.

In summary, HE significantly mitigates privacy risks caused by inference attacks directly using intermediate results and gradients exchanged between different parties. HE is often used along with other defense strategies, such as DO and SS defined in Section \ref{ssec:defense}.
However, they could still be vulnerable to passive and active insider attackers even without compromising HE key pairs. A passive attacker always has access to decrypted local gradients to update local models and uses them solely to infer private data, as demonstrated by Gradient Matching-based label inference attacks \cite{liu2021batch,zou2022defending} and Gradient Linear Inversion attacks \cite{hu2022vertical} discussed in Section \ref{sss:labelgradientattack} and \ref{sss:gradfeatureattacks}. An active attacker can maliciously modify the computation under HE to derive private feature information, as demonstrated by the reverse sum attack \cite{weng2020privacy} against SecureBoost in Section \ref{ssec:resdataintraining}.

\noindent\textbf{Functional Encryption (FE) based defense}\quad  Xu et al.~\cite{xu2021fedv} proposed FedV, a VFL framework that employs functional encryption (FE) for secure gradient computation for linear/logistic and SVM Models. It assumes an aggregator that acts as a semi-honest coordinator interacting with the active party and passive parties. In addtion, it requires a trusted third-party authority (TPA) crypto-infrastructure to enable functional encryption for secure aggregation. FedV uses a Multi-Input Functional Encryption (MIFE) scheme~\cite{abdalla2018multi} for the inner product functionality for Feature Dimension Secure Aggregation, and a Single-Input Functional Encryption (SIFE) scheme~\cite{abdalla2015simple} for Sample Dimension Secure Aggregation.
However, it requires the active party to share labels with the aggregator in plaintext for non-linear models, and reveals the intermative results to the aggregator. To address the information leakage in FedV, Chen et al.~\cite{chen2023quadratic} proposed SFedV, a privacy enahancement with Multi-input Quadratic Functional Encryption qMIFE~\cite{agrawal2022multi}.

\noindent\textbf{Secure Multi-party Computation (MPC) based defense}\quad  Similarly to the previous section, we discuss MPC based defense mechanisms by the types of VFL models. They employ various MPC techniques including SS(Secret Sharing), OT(Oblivious Transfer) and GC(Garbled Circuit), which are frequently integrated with HE.    

\indent\textit{VFL linear/logistic regression:}\quad
Gascon et al. \cite{gascon2016secure} proposed a secure protocol for VFL linear regression model, which combines the use of Yao’s GC, SS and OT. It consists of two phases: the aggregation phase that combines data from different parties and the phase of solving the system of linear equations. The aggregation phase uses OT and SS techniques to securely compute the inner product of the vectors of two parties, and each party obtains a secret share of the result. The solving phase employs two non-colluding servers, Crypto Server Provider and Evaluator,  and solves distributed linear systems using gradient descent within a GC protocol. An improvement is proposed by Giacomelli et al. \cite{giacomelli2018privacy} to reduce the communication overhead by only employing linearly-homomorphic encryption instead of Yao's GC.

The SecureML work proposed by Mohassel et al. \cite{mohassel2017secureml} presents secure protocols against semi-honest adversaries for privacy-preserving machine learning using SGD, including linear regression, logistic regression, and neural networks. It assumes that secret shares of participants' private data are distributed among two non-colluding servers with an arbitrary dataset partitioning (horizontal or vertical). The protocols address the performance issue by vectorization in the shared setting and a data-independent offline phase that generates multiplication triplets based on linearly HE and OT for secure arithmetic operations on shared decimal numbers. SecureML shows a significant speedup in the linear regression model compared to the previous work by Gascon et al. \cite{gascon2016secure}. Compared to SecureML, ABY$^3$ proposed by Mohassel et al.~\cite{Mohassel2018aby3} implements a three-server model. It introduces a framework for efficient conversion between the arithmetic, binary, and Yao sharing types, with security against malicious adversaries.

Chen et al. \cite{chen2021homomorphic} proposed CAESAR, a secure VFL logistic regression that addresses the efficiency problem when dealing with high-dimensional sparse features. By applying SS in HE field, CAESAR introduces a secure sparse matrix multiplication protocol between two parties, where each party finally obtains a secret share of the result, without exposing their private features and labels. Based on this protocol, CAESAR ensures that models are always secret shares during model training, and each party reconstructs its local model after training.
Fan et al. \cite{fan2023vertices} proposed an SS-based approach using a trusted coordinator, called VERTICES, for two-party VFL linear and logistc regression. VERTICES uses the trusted coordinator for secret multiplication to remove the dependency on multiplication triplet computation and its cost. It also applies random masks within the SS protocol to reduce the communication and computation cost. Zhou et al. \cite{zhou2021privacy} proposed a general secure aggregation protocol PFLF that combines SS and HE for privacy-preserving aggregation between an active party and passive parties. Using this protocol, the authors developed secure multiparty entity matching and logistic regression algorithms in VFL. However, it aims only to prevent the active party from inferring the local data of passive parties and does not consider label privacy. The training algorithm discloses the final results of the aggregation and the sample errors to all passive parties.

\indent\textit{VFL Decision Tree:}\quad
Wu et al. \cite{wu2020privacy} introduced a privacy-preserving method called Pivot to safeguard the privacy of decision trees in VFL against a semi-honest adversary that may compromise $m-1$ out of $m$ clients.
Assuming that one party holds the labels, Pivot combines partially homomorphic encryption (PHE) and additive SS scheme SPDZ~\cite{damgaard2012multiparty} that supports secure approximate division and comparison. Its training process produces secretly shared best split without disclosing intermediate results. To further hide prediction labels on leaf nodes and split thresholds on internal nodes of a tree model from clients, Pivot uses private information retrieval(PIR)~\cite{wu2018privacy} to privately select the split during training, and converts the encrypted split thresholds and leaf labels to secret shares among clients after training. In addition, the authors further extended Pivot with zero-knowledge proofs(ZKP) and authenticated shares to defend against malicious attackers, and with differential privacy to protect individual privacy.

Fang et al. \cite{fang2021large} proposed an SS based gradient tree boosting (SS-XGB) framework for two-party VFL. This framework also uses SS based arithmetic and argmax operations as \cite{wu2020privacy}. In addition, a secure permutation protocol that permutes a secret shared vector with a given rank is employed to sort the first-order and second-order gradients for computing the sum of gradients in XGB. It improves the communication cost incurred by matrix multiplication operations in SS-XGB. Based on the implementation of secure permutation, two variants are introduced, in which HEP-XGB uses HE based permutation and CRP-XGB uses CR based permutation that utilizes correlated randomness(CR) in Beaver’s multiplication triplet.
The study by Xie et al. \cite{xie2022efficient} also addressed high computation and communication overheads associated with SS-based solutions~\cite{wu2020privacy}. They proposed a secure multi-party federated XGBoost learning framework called MP-FedXGB. This framework operates under the assumption of a semi-honest third-party coordinator and is not limited to two-party VFL. Its efficiency is primarily attributed to the reshaping of the calculation process of the split criterion and leaf node weight calculation, which removes the division operations under SS in the training process. Jin et al.~\cite{jin2022towards} proposed a VFL XGBoost system, CryptoBoost, which implements Multi-Party Homomorphic Encryption (MHE) based secure protocols for division, exponent, comparison and sorting in training.
Lu et al .\cite{Lu2023Squirrel} proposed a two-party VFL training framework named Squirrel for Gradient Boosting Decision Tree (GBDT), offering performance improvements over previous works like Pivot ~\cite{wu2020privacy} and HEP-XGB~\cite{fang2021large}. The key to Squirrel's efficiency lies in its integration with the Ferret OT protocol~\cite{yang2020ferret} to keep sample indicator vector at tree nodes private, lattice-based HE for gradient aggregation and two-party private protocol for computing sigmoid function. Zheng et al . proposed \cite{zheng2023privet} Privet, a privacy-preserving VFL framework for gradient-boosted decision tables. Similar to SS-XGB~\cite{fang2021large}, it uses SS based arithmetic and CR based permutation to implement secure training of a distributed decision table.

\indent\textit{VFL Neural Network:}
Fu et al. \cite{fu2022blindfl} proposed a VFL framework BlindFL that introduces secure federated source layers to unite different input features of each participant without disclosing private data. The output of federated source layers from each party is further aggregated as the input of the top model for classification at the active party. By combining HE and SS, BlindFL securely trains federated source layers with privacy-preserving matrix multiplication and embedding lookup, without disclosing the intermediate results and final model weights to each party. The work SFTL by Liu et al. ~\cite{liu2020secure} introduced a SS-based protocol to encrypt information exchanged between two parties but the local models are in plain text and the prediction phase leaks the labels of samples from one party to another party. Sharma et al. \cite{sharma2019secure} proposed an improvement to the efficiency of SFTL by using the ABY framework~\cite{demmler2015aby} for the semi-honest setting and the SPDZ framework~\cite{damgaard2012multiparty} for the malicious setting. Yang et al.~\cite{yang2022hybrid} proposed a hybrid secure protocol for two-party VFL which adopts arithmetic SS for linear operations and Yao's GC for non-linear operations in neural networks. Li et al. \cite
{li2023fedvs} proposed a split VFL framework, FedVS, to simultaneously address straggling clients in training and data/model inference attacks using clients' output embeddings. To prevent privacy leakage from the clients' uploaded embeddings, it employs an SS-based approach that uses Lagrange coded computing (LCC) \cite{yu2019lagrange} for sharing multiple secrets. Each client uses SS to distribute its training data and its local polynomial network model~\cite{livni2014computational} to guarantee lossless reconstruction of the sum of the embeddings of the clients. 

\begin{table}[]
\scriptsize
\caption{Summary of existing cryptographic defenses in model training phase of VFL.}\label{tab:cryptodefenseintraining}
\begin{tabular}{|c|c|c|c|c|c|cc|} 

\hline
\multirow{2}{*}{\begin{tabular}[c]{@{}c@{}}Existing\\ Crypto-Defenses\end{tabular}} & \multirow{2}{*}{\begin{tabular}[c]{@{}c@{}}Defense\\ Strategy\end{tabular}} & \multirow{2}{*}{Model} & \multirow{2}{*}{\# parties} & \multirow{2}{*}{Coordinator} & \multirow{2}{*}{Security} & \multicolumn{2}{c|}{Threat Model}              \\ \cline{7-8} 
                                                                                    &                                                                             &                        &                            &   &                           & \multicolumn{1}{c|}{Attacker} & collusion \\ \hline
\cite{yang2019federated}                                                            & HE                                                                          & LR                     & 2                          & Y       &   S-1                  & \multicolumn{1}{c|}{P}             & N         \\ \hline
HardyLR\cite{hardy2017private}                                                      & HE                                                                          & LR                     & 2                          & Y       &   S-1                  & \multicolumn{1}{c|}{P}             & N         \\ \hline
\cite{giacomelli2018privacy}                                                        & HE                                                                          & LR                     & $2^*$                      & Y      &    S-3                  & \multicolumn{1}{c|}{P}             & N         \\ \hline
BaiduLR\cite{yang2019parallel}                                                      & HE                                                                          & LR                     & 2                          & N        &  S-1                   & \multicolumn{1}{c|}{P}             & N         \\ \hline
\cite{yang2019quasi}                                                                & HE                                                                          & LR                     & 2                          & Y        &  S-1                & \multicolumn{1}{c|}{P}             & N         \\ \hline
SecureLR\cite{he2021secure}                                                         & HE                                                                          & LR                     & $\ge 2$                    & N       &    $a:$S-1, $p:$S-2              & \multicolumn{1}{c|}{P}             & N         \\ \hline
VPRF\cite{hou2021verifiable}                                                        & HE                                                                          & Tree                   & $\ge 2$                    & Y       &   S-1                  & \multicolumn{1}{c|}{P}             & Y         \\ \hline
RevFRF\cite{liu2021revfrf}                                                          & HE                                                                          & Tree                   & $\ge 2$                    & Y        &  S-1                  & \multicolumn{1}{c|}{P}             & Y         \\ \hline
VF-CART\cite{XU2023VFCART}                                                          & HE                                                                          & Tree                   & $\ge 2$                    & Y        &  S-1                  & \multicolumn{1}{c|}{P}             & Y         \\ \hline
SecureGBM\cite{feng2019securegbm}                                                   & HE                                                                          & Tree                   & 2                          & N      &    S-1                  & \multicolumn{1}{c|}{P}             & N         \\ \hline
SecureBoost\cite{cheng2021secureboost}                                              & HE                                                                          & Tree                   & $\ge 2$                    & N       &    S-1                & \multicolumn{1}{c|}{P}             & N         \\ \hline
SecureBoost+\cite{chen2021secureboost+}                                             & HE                                                                          & Tree                   & $\ge 2$                    & N        &   S-1               & \multicolumn{1}{c|}{P}             & N         \\ \hline
Vf2boost\cite{fu2021vf2boost}                                                       & HE                                                                          & Tree                   & $\ge 2$                    & N       &    S-1                & \multicolumn{1}{c|}{P}             & N         \\ \hline
\cite{chamani2020mitigating}                                                        & HE(+TEE)                                                                          & Tree                   & $\ge 2$                    & N        &   $a:$S-2, $p:$S-3                 & \multicolumn{1}{c|}{P}             & N         \\ \hline
ACML\cite{zhang2020additively}                                                      & HE                                                                          & NN                     & 2                          & N       &     S-1                & \multicolumn{1}{c|}{P}             & N         \\ \hline
HFTL\cite{liu2020secure}                                                            & HE                                                                          & NN                     & 2                          & N        &     S-1                & \multicolumn{1}{c|}{P}             & N         \\ \hline
PrADA\cite{kang2022privacy}                                                         & HE                                                                          & NN                     & $\ge 2$                    & N        &     S-1               & \multicolumn{1}{c|}{P}             & N         \\ \hline
Fedsgc\cite{cheung2021fedsgc}                                                       & HE                                                                          & GNN                    & 2                          & N        &     S-1                & \multicolumn{1}{c|}{P}             & N         \\ \hline
SecureSL\cite{yang2023securesl}                                                     & HE                                                                          & NN                     & $\ge 2$                    & N       &      S-3              & \multicolumn{1}{c|}{P}             & Y         \\ \hline
SplitWays\cite{khan2023split}                                                       & HE                                                                          & NN                     & 2                          & N        &      S-1                & \multicolumn{1}{c|}{P}             & N         \\ \hline
FedV\cite{xu2021fedv}                                                               & FE                                                                          & LR                     & $\ge 2$                    & Y        &     S-1               & \multicolumn{1}{c|}{P}             & Y         \\ \hline
SFedV\cite{chen2023quadratic}                                                       & FE                                                                          & LR                     & $\ge 2$                    & Y        &     S-3               & \multicolumn{1}{c|}{P}             & Y         \\ \hline
GasconLR\cite{gascon2016secure}                                                             & MPC (GC+SS+OT)                                                              & LR                     & $\ge 2$                    & Y        &  S-2                    & \multicolumn{1}{c|}{P}             & N         \\ \hline
SecureML\cite{mohassel2017secureml}                                                 & MPC (HE+SS+OT)                                                              & LR, NN                 & $2^*$                      & Y        &       S-3             & \multicolumn{1}{c|}{P}             & N         \\ \hline
ABY$^3$\cite{Mohassel2018aby3}                                                      & MPC (GC+SS+OT)                                                              & LR, NN                 & $3^*$                      & Y       &        S-3             & \multicolumn{1}{c|}{A}             & N         \\ \hline
CAESAR\cite{chen2021homomorphic}                                                    & MPC (HE+SS)                                                                 & LR                     & $\ge 2$                    & Y      &         S-2              & \multicolumn{1}{c|}{P}             & N         \\ \hline
VERTICES \cite{fan2023vertices}                                                     & MPC (SS)                                                                    & LR                     & 2                          & Y       &        S-2              & \multicolumn{1}{c|}{P}             & N         \\ \hline
PFLF\cite{zhou2021privacy}                                                          & MPC(HE+SS)                                                                  & LR                     & $\ge 2$                    & N        &    a:S-1 p:S-2                 & \multicolumn{1}{c|}{P}             & Y         \\ \hline
Pivot(basic)\cite{wu2020privacy}                                                           & MPC(HE+SS)                                                                  & Tree                   & $\ge 2$                    & N        &        S-2             & \multicolumn{1}{c|}{P}             & Y         \\ \hline
Pivot\cite{wu2020privacy}                                                           & MPC(HE+SS)                                                                  & Tree                   & $\ge 2$                    & N        &        S-3             & \multicolumn{1}{c|}{P}             & Y         \\ \hline
Pivot(extension)\cite{wu2020privacy}                                                           & MPC(HE+SS)+ZKP                                                                  & Tree                   & $\ge 2$                    & N       &    S-3          & \multicolumn{1}{c|}{A}             & Y         \\ \hline
SS-XGB\cite{fang2021large}                                                          & MPC(SS)                                                                     & Tree                   & 2                          & Y         &     S-2               & \multicolumn{1}{c|}{P}             & N         \\ \hline
HEP-XGB\cite{fang2021large}                                                         & MPC(HE+SS)                                                                  & Tree                   & 2                          & Y         &     S-2               & \multicolumn{1}{c|}{P}             & N         \\ \hline
CRP-XGB\cite{fang2021large}                                                         & MPC(SS+DO)                                                                  & Tree                   & 2                          & Y         &    S-2                & \multicolumn{1}{c|}{P}             & N         \\ \hline
MP-FedXGB\cite{xie2022efficient}                                                    & MPC(SS)                                                                     & Tree                   & $\ge 2$                    & Y         &     S-2         & \multicolumn{1}{c|}{P}             & N         \\ \hline
CryptoBoost\cite{jin2022towards}                                                    & MPC(HE+SS)                                                                  & Tree                   & $\ge 2$                    & N         &     S-2           & \multicolumn{1}{c|}{P}             & Y         \\ \hline
Squirrel\cite{Lu2023Squirrel}                                                       & MPC(HE+SS+OT)                                                               & Tree                   & 2                          & N         &     S-2               & \multicolumn{1}{c|}{P}             & N         \\ \hline
Privet\cite{zheng2023privet}                                                        & MPC(SS)                                                                     & Tree                   & $\ge 2$                    & N         &      S-2              & \multicolumn{1}{c|}{P}             & Y         \\ \hline
BlindFL\cite{fu2022blindfl}                                                         & MPC(HE+SS)                                                                  & NN                     & $\ge 2$                    & N         &     a:S-1, p:S-3              & \multicolumn{1}{c|}{P}             & N         \\ \hline
SFTL\cite{liu2020secure}                                                            & MPC(SS)                                                                     & NN                     & 2                          & N         &      S-1              & \multicolumn{1}{c|}{P}             & N         \\ \hline
SS+ABY\cite{sharma2019secure}                                                       & MPC(SS+OT)                                                                  & NN                     & 2                          & N         &      S-1              & \multicolumn{1}{c|}{P}             & N         \\ \hline
SPDZ\cite{sharma2019secure}                                                         & MPC(SS)                                                                     & NN                     & $\ge 2$                    & N         &      S-1              & \multicolumn{1}{c|}{A}             & N         \\ \hline
\cite{yang2022hybrid}                                                               & MPC(SS+GC)                                                                  & NN                     & 2                          & N         &     S-1              & \multicolumn{1}{c|}{P}             & N         \\ \hline
FedVS\cite{li2023fedvs}                                                             & MPC(SS)                                                                     & NN                     & $\ge 2$                    & N         &     S-1               & \multicolumn{1}{c|}{P}             & Y         \\ \hline
\end{tabular}
\end{table}

Table \ref{tab:cryptodefenseintraining} summarizes all the cryptographic defenses that we reviewed. Note that $2^*$ and $3^*$ represent special scenarios where 2-server and 3-server architectures are used for privacy-preserving machine learning. For VFL, they involve the outsourcing of data from data owners to the servers to implement VFL. The `Coordinator' column indicates whether a third-party coordinator server is used in the defense mechanism. The `Security' column specifies the security level of a defense mechanism, with `a' representing the active party and `p' the passive party. In the `Attacker type' column, `P' represents a passive attacker and `A' an active attacker. The `Collusion' column indicates whether the defense mechanism addresses colluding adversaries.

\vspace{5pt}
\subsubsection{Non-cryptographic Defense}~

\noindent\textbf{Obfuscation Based Defense}
This type of defense uses obfuscated gradients, labels, or intermediate results to counter the feature/label inference attack.

\indent\textit{1. Obfuscated Features.} 
Gu et al. \cite{gu2023fedpass} proposed FedPass, a privacy-preserving VFL framework that adopts adaptive obfuscation in both the active and passive parties, by inserting passports into their models to protect data features and labels. The adaptive obfuscation functions act on private features and intermediate results and are optimized during the learning stage. 
Gu et al.~\cite{gu2020federated} proposed a federated doubly stochastic kernel learning algorithm (FDSKL) for VFL. To avoid directly transferring the local data for training, FDSKL applies a random linear transformation on the data of each client, and uses a tree structure-based communication scheme for the result aggregation to prevent the leakage of information used by the linear transformation.

\indent\textit{2. Obfuscated Gradients.} To address the feature inference attack, the authors of the CAFE attack\cite{NEURIPS2021_08040837} proposed a fake gradient approach. It generates fake gradients from a normal distribution and matches them with true gradients by $l_2$ norm to find the closest fake gradient for each true gradient. The local workers sort these fake gradients to the same order of corresponding true gradients and upload them to the server, in order to achieve similar learning performance to the true gradients. The experiment shows that it effectively mitigates the data leakage of CAFE.

To address the label leakage in two-party split learning from gradients, Li et al.\cite{li2021label} proposed a heuristic approach \textit{max\_norm}. It adds zero-mean Gaussian noise with non-isotropic and example-dependent covariance to the gradients, while keeping the similar direction and making the squared 2-norm of each perturbed gradient close to the largest one in the same batch.  

Because \textit{max\_norm} does not have theoretical justification, Li et al.\cite{li2021label} further proposed another method named Marvell that aims to protect against the entire class of scoring-based attacks: Gradient Norm Scoring-Based Inference and Gradient Direction Scoring-Based Inference, which we covered in Section \ref{sss:labelgradientattack}. Marvell applies noise perturbation through an optimal noise structure that minimizes the worst-case adversarial scoring functions' leak AUC metric under a utility constraint. The noise structure consists of two zero-mean different random noise distributions that are used to perturb positive and negative gradients, respectively.

In \cite{fu2022label} Fu et al. evaluated four different defense approaches against the DLI attack including noisy gradients, gradient compression, privacy-preserving deep learning (PPDL)~\cite{shokri2015privacy}, and DiscreteSGD. The DiscreteSGD method discretizes gradient values by categorizing them into distinct bins. The experiment results showed that, while noisy gradients and PPDL effectively mitigated the attack, the other two approaches did not show similar success.

\indent\textit{3. Obfuscated Labels.}
To defend against the Gradient Matching-based batch label inference attack, Liu and Zou et al. \cite{liu2021batch, zou2022defending} proposed a defense mechanism that uses an autoencoder and entropy regularization to disguise the true labels. The active party learns a confusional autoencoder (CoAE) to transform an original label to a soft label with higher probability for each alternative class. After training, the active party can leverage the CoAE to produce and use fake labels labels to compute gradients for training, thereby mitigating label leakage from gradients. During inference time, the active party transforms the predicted labels back to true labels with CoAE decoder.
Compared to previous methods such as gradient noise \cite{xin2020private} and gradient sparsification \cite{han2020adaptive,mitra2021linear}, CAE addresses the significant loss in task accuracy caused by the tight coupling between label accuracy and the accuracy of the main task\cite{liu2021defending}. Furthermore, an enhanced version of CAE, called Discrete-label-enhanced CAE (DCAE), is introduced to enhance robustness against attackers with prior knowledge. DCAE demonstrates sufficient resilience when facing attackers who possess prior knowledge.

\indent\textit{4. Obfuscated Intermediate Data.}
To prevent leakage of label privacy, Zhang et al.~\cite{zhang2021secure} proposed a backward updating mechanism that avoids directly using label information to compute gradients. It applies random masking on the intermediate results and adopts the same approach as FDSKL for secure aggregation. Wang et al.~\cite{wang2022feverless} proposed a secure protocol, FEVERLESS, to train XGBoost for VFL where training labels are distributed over multiple clients. FEVERLESS protects the privacy of the features and labels from inference attacks during aggregation by applying random masking to gradients and hessians sent from clients. Random masking is generated by key derivation with Diffie-Hellman key exchange for canceling out. In addition, it applies differential privacy to aggregated values to prevent gradient leakage. Qiu et al.~\cite{qiu2023vfedsec} proposed vFedSec, a VFL framework that requires a central server and employs a secure layer at the server to aggregate the intermediate output of each client. It uses pairwise shared secrets among clients to enable each client to add random noise to the intermediate output with the sum of added noise among clients being zero. The gradients of the local models are masked in the same way and sent to the server to compute the average used to update the local models.

\noindent\textbf{Adversarial training (AT) based defense} \quad Existing works that employ this defense strategy are categorized into two groups based on their protection goals: feature privacy and label privacy. Overall, these defense methods share a similar approach, involving the introduction of additional optimization objectives aimed at reducing the correlation between features/labels and intermediate results to reduce privacy leakage.

\textit{Feature Privacy:}\quad
Vepakomma et al. \cite{vepakomma2020nopeek} proposed a method NoPeek to defend against feature reconstruction attacks for split learning. It is designed to reduce the correlation between activations from protected cut layers and raw features while preserving distance correlation with labels, which is achieved by adding the correlation to the classification loss function to optimize. Additionally, the method starts with a local de-correlation step at the client to avoid information leakage during the initial training iterations.
Sun et al. \cite{sun2021defending} proposed a framework called DRAVL to defend feature reconstruction attacks against passive parties. It includes three modules: adversarial reconstructor (AR), noise regularization (NR) and distance correlation (DC). The loss function involves three additional loss terms that correspond to each module, respectively. DRAVL effectively enlarges the reconstruction error by maximizing the reconstruction loss by using a gradient reversal layer in the model and minimizing the distance between reconstructed output and random Gaussian noise as well as the correlation between raw data and activation outputs from passive parties.
 Zhang and Turina et al. \cite{turina2021federated, zhang2023privacy} proposed a client-based approach to protect the privacy of client data for federated split learning. It uses an idea similar to NoPeek, which mitigates information leakage by minimizing the distance correlation between input features and intermediate data. Li et al.~\cite{li2022ressfl} proposed a split federated learning framework ResSFL to defend against the intermediate result inversion attack that reconstructs a client's data from its model activation output. The attacker-aware training process trains a feature extractor with strong resistance against inversion attack at the client side, by using an inversion score that measures the quality of data reconstruction in the objective function.

\textit{Label Privacy:}\quad
Sun et al. \cite{sun2022label} introduced a label inference attack using intermediate embedding from cut layers and proposed a defense method. The defense uses an additional optimization target in the loss function that is to minimize the distance correlation between intermediate embeddings and private labels during the training.
To address the instance space-based label inference attack against Tree VFL in Section \ref{sss:labelmedattack}, Takahashi et al. \cite{takahashi2023eliminating} proposed a defense mechanism named ID-LMID, based on mutual information regularization. ID-LMID prevents label leakage by reducing mutual(MI) information between the label and instance space in the split finding computation. The active party reduces MI by restricting the choice of split candidates from a passive party and the broadcast of certain instance spaces.
Han et al. \cite{han2023gan} proposed a GAN based method to mitigate label pivacy leakage in split learning. The active party trains a GAN model using the loss function that consists of the distance between the distribution of prediction label and true label, as well as an additional cross-entropy (CE) loss between the intermediate result from passive parties and a randomized label derived from true label. This approach generates more mixed intermediate gradients compared to vanilla splitNN, thereby mitigating the label privacy leakage from gradients.

\noindent\textbf{Differential Privacy (DP) based defense} \quad We discuss this defense strategy from the type of models to which differential privacy is applied, as the design of differential privacy algorithms can vary significantly depending on the type of model.

\textit{VFL Clustering:}\quad
Li et al. \cite{li2022differentially} proposed a differentially private vertical federated k-means clustering. To protect local data in a DP manner, the basic idea is to generate a differentially private global approximation of the data points and perform k-means clustering on it.  In the algorithm, each party uses differentially private algorithms to compute the local centers and encode the membership information of each local cluster. The central server builds a weighted grid in which the grid nodes are the Cartesian product of the local centers of different parties and runs a central k-means algorithm on the grid nodes to generate the final k centers. The DP local clustering algorithm is adapted from DPLSF~\cite{charikar2002similarity,dpcluster}, and the membership encoding and weight estimation are based on DP Flajolet-Martin (FM) sketch~\cite{smith2020flajolet}.

\textit{VFL LR:}\quad
Chen et al. \cite{chen2020vafl} proposed an asynchronous VFL framework for both logistic regression(LR) and neural network models that eliminates the requirement of coordination between clients. To achieve differential privacy on private feature data, each client applies the Gaussian mechanism to perturb its local embedding output. Wang et al. \cite{wang2020hybrid} introduced a differentially private framework for VFL to learn a generalized linear model. The authors analyze the sensitivity of intermediate result (IR) exchanged between passive parties and single active party, and perturb IR with the Gaussian mechanism to achieve DP w.r.t. IR and joint differential privacy \cite{kearns2014mechanism} w.r.t. model weights. To address Gradient Linear Inversion attacks against label and feature privacy in vertical LR, Hu et al.\cite{hu2022vertical} proposed a countermeasure that applies the Gaussian mechanism to mask the information exchanged between the passive party and the active party in the HE training protocol. It aims to prevent both parties from learning the precise values used for inference attacks after solving the linear system. To address the residual variable-based label inference attack in vertical LR that we discussed in Section \ref{sss:labelgradientattack}, Tan et al. \cite{tan2022residue} proposed to perturb residues with random noise from a Laplace distribution in an additive or multiplicative way to achieve $\epsilon$-LDP. Furthermore, they proposed a hybrid mechanism that combines HE with random response~\cite{wang2016using} mechanism to perturb the indicator vector in the training protocol to achieve $\epsilon$-LDP.

\textit{VFL Tree:}\quad
Tian et al. \cite{tian2020federboost} proposed a FederBoost framework for GBDT in VFL that satisfies $\epsilon$-local element-level DP defined in the paper. In FederBoost, the active party collects differentially private ordering information of sample features from passive parties to train GBDT. Each passive party partitions a feature into buckets and samples are assigned to each bucket in a differentially private way similar to the randomized response mechanism~\cite{wang2016using} for local differential privacy (LDP). 
Li et al \cite{li2022opboost} proposed OpBoost, a VFL tree boosting framework that shares the same idea as FederBoost. Its privacy-preserving order-preserving algorithm discretizes feature values to partitions and assigns samples to partitions using the exponential mechanism, which satisfies distance-based LDP. Zhu et al. \cite{zhu2021pivodl} developed a secure learning system named PIVODL to privately train the XGBoost decision tree in VFL with labels distributed on multiple clients. PIVODL uses Paillier HE to encrypt the information exchanged between source clients and split clients to determine the optimal node split. To further prevent label leakage to the source clients caused by leaf weights during prediction update, PIVODL applies the Gaussian mechanism to the clipped weight values at split clients to achieve differential privacy. To address the instance space-based label inference attack against Tree VFL in Section \ref{sss:labelmedattack}, Takahashi et al. \cite{takahashi2023eliminating} proposed a defense mechanism Grafting-LDP. It adds noise to labels to satisfy $\epsilon$-Label DP and applies post-processing at the active party to fix the problematic split with the feature of the active party using clean labels. However, Grafting-LDP cannot be applied to boosting methods like XGBoost since each tree is not independently trained and post-processing on the first tree requires to retrain the rest of trees.

\textit{VFL NN:}\quad
Chen et al. \cite{chen2020vafl} applies the Guassian mechanism to perturb the output of the local NN model to protect the private feature of each client. Yang et al. \cite{yang2022differentially} proposed a split learning framework named TPSL to protect label privacy with transcript differential privacy which guarantees that the messages exchanged between parties satisfy $(\epsilon,\delta)$-DP.
TPSL perturbs the gradients by adding noise from the Laplace distribution and Bernoulli distribution to the gradients only in the optimal direction and results in $(\epsilon, 0)$-transcript DP.
Ranbaduge et al. \cite{ranbaduge2022differentially} analyzed the sensitivity of the gradient estimate for applying the Gaussian mechanism to VFL training and empirically investigated the VFL model accuracy under various DP privacy budgets.
Chen et al. \cite{chen2020vertically} proposed a privacy-preserving approach VFGNN for training graph neural network models in VFL, where participants have the same nodes but different features and edges. VFGNN protects local data by applying differential privacy to local node embeddings from participants. Oh et al. \cite{oh2022differentially} proposed DP-CutMixSL, a differentially private (DP) split learning framework for the vision transformer. It applies the Gaussian mechanism to the cutout smashed data from clients. The authors also theoretically prove that the proposed randomized CutMix operation for inter-dataset interpolations during the training amplifies the DP guarantee of smashed data. To address the feature inference attack, Mao et al. \cite{mao2022secure} introduced a new activation function Randomized-Response ReLU (R3eLU) for Split neural networks, which consists of a randomized response procedure and a Laplace mechanism. With R3eLU, the training process perturbs information propagation in both forward and backward directions and achieves $\epsilon$-differential privacy. The authors also proposed to dynamically allocate the privacy budget to different features based on their importance and to different iterations for better budget utilization. Tran et al. \cite{tran2023privacy} proposed PBM-VFL, a VFL training algorithm with differential privacy to protect the private features. In this approach, passive parties use the Poisson Binomial Mechanism \cite{chen2022poisson} to produce quantized embeddings from their local models, ensuring differential privacy for the output.  These embeddings are then securely aggregated and fed into the active party's top model. The aggregation is performed using an MPC protocol to ensure the confidentiality of the embeddings.

Table \ref{tab:noncryptodefenseintraining} summarizes all the noncryptographic defenses that we reviewed. Compared to Table \ref{tab:cryptodefenseintraining}, an additional `Privacy' column is included to indicate what type of inference attacks a defense mechanism aims to counter, whether it be feature or label inference attacks.  

\begin{table}[]
\scriptsize
\caption{Summary of existing Non-cryptographic defenses in model training phase of VFL.}\label{tab:noncryptodefenseintraining}
\begin{tabular}{|c|c|c|c|cc|cc|}
\hline
\multirow{2}{*}{\begin{tabular}[c]{@{}c@{}} NonCrypto \\Defenses\end{tabular}} & \multirow{2}{*}{\begin{tabular}[c]{@{}c@{}}Defense\\ Strategy\end{tabular}} & \multirow{2}{*}{Model} & \multirow{2}{*}{\# parties} & \multicolumn{2}{c|}{Threat Model}              & \multicolumn{2}{c|}{Privacy}                     \\ \cline{5-8} 
                                                                                    &                                                                             &                        &                             & \multicolumn{1}{c|}{Attacker type} & Collusion & \multicolumn{1}{c|}{Feature}      & Label        \\ \hline
FedPass\cite{gu2023fedpass}                                                         & DO(Feature,Intermediate)                                                    & NN                     & $\ge 2$                     & \multicolumn{1}{c|}{P}             & N         & \multicolumn{1}{c|}{$\checkmark$} & $\checkmark$ \\ \hline
FDSKL\cite{gu2020federated}                                                         & DO(Feature,Intermediate)                                                    & Kernel                 & $\ge 2$                     & \multicolumn{1}{c|}{P}             & N         & \multicolumn{1}{c|}{$\checkmark$} & $\checkmark$ \\ \hline
CAFE\cite{NEURIPS2021_08040837}                                                     & DO(Gradient)                                                                & NN                     & $\ge 2$                     & \multicolumn{1}{c|}{P}             & N         & \multicolumn{1}{c|}{$\checkmark$} &              \\ \hline
\textit{max\_norm}\cite{li2021label}                                                & DO(Gradient)                                                                & NN                     & 2                           & \multicolumn{1}{c|}{P}             & N         & \multicolumn{1}{c|}{}             & $\checkmark$ \\ \hline
Marvell\cite{li2021label}                                                           & DO(Gradient)                                                                & NN                     & 2                           & \multicolumn{1}{c|}{P}             & N         & \multicolumn{1}{c|}{}             & $\checkmark$ \\ \hline
CoAE\cite{liu2021batch,zou2022defending}                                                          & DO(Label)                                                                   & LR,NN                  & $\ge 2$                     & \multicolumn{1}{c|}{P}             & N         & \multicolumn{1}{c|}{}             & $\checkmark$ \\ \hline
\cite{zhang2021secure}                                                              & DO(Intermediate)                                                            & LR                     & $\ge 2$                     & \multicolumn{1}{c|}{P}             & Y         & \multicolumn{1}{c|}{}             & $\checkmark$ \\ \hline
FEVERLESS\cite{wang2022feverless}                                                   & DO(Intermediate) + DP                                                       & Tree                   & $\ge 2$                     & \multicolumn{1}{c|}{P}             & Y         & \multicolumn{1}{c|}{$\checkmark$} & $\checkmark$ \\ \hline
vFedSec\cite{qiu2023vfedsec}                                                        & DO(Intermediate)                                                            & NN                     & $\ge 2$                     & \multicolumn{1}{c|}{P}             & Y         & \multicolumn{1}{c|}{$\checkmark$} & $\checkmark$ \\ \hline
NoPeek\cite{vepakomma2020nopeek}                                                    & AT                                                                          & NN                     & $\ge 2$                     & \multicolumn{1}{c|}{P}             & N         & \multicolumn{1}{c|}{$\checkmark$} &              \\ \hline
DRAVL\cite{sun2021defending}                                                        & AT                                                                          & NN                     & 2                           & \multicolumn{1}{c|}{P}             & N         & \multicolumn{1}{c|}{$\checkmark$} &              \\ \hline
\cite{turina2021federated,   zhang2023privacy}                                      & AT                                                                          & NN                     & $\ge 2$                     & \multicolumn{1}{c|}{P}             & N         & \multicolumn{1}{c|}{$\checkmark$} &              \\ \hline
ResSFL\cite{li2022ressfl}                                                           & AT                                                                          & NN                     & $\ge 2$                     & \multicolumn{1}{c|}{P}             & N         & \multicolumn{1}{c|}{$\checkmark$} &              \\ \hline
\cite{sun2022label}                                                                 & AT                                                                          & NN                     & 2                           & \multicolumn{1}{c|}{P}             & N         & \multicolumn{1}{c|}{}             & $\checkmark$ \\ \hline
ID-LMID\cite{takahashi2023eliminating}                                              & AT                                                                          & Tree                   & $\ge 2$                     & \multicolumn{1}{c|}{P}             & N         & \multicolumn{1}{c|}{}             & $\checkmark$ \\ \hline
\cite{han2023gan}                                                                   & AT                                                                          & NN                     & $\ge 2$                     & \multicolumn{1}{c|}{P}             & N         & \multicolumn{1}{c|}{}             & $\checkmark$ \\ \hline
\cite{li2022differentially}                                                         & DP                                                                          & Clustering             & $\ge 2$                     & \multicolumn{1}{c|}{P}             & N         & \multicolumn{1}{c|}{$\checkmark$} &              \\ \hline
VAFL\cite{chen2020vafl}                                                             & DP                                                                          & LR,NN                  & $\ge 2$                     & \multicolumn{1}{c|}{P}             & N         & \multicolumn{1}{c|}{$\checkmark$} &              \\ \hline
\cite{wang2020hybrid}                                                               & DP                                                                          & LR                     & $\ge 2$                     & \multicolumn{1}{c|}{P}             & N         & \multicolumn{1}{c|}{$\checkmark$} & $\checkmark$ \\ \hline
\cite{hu2022vertical}                                                               & DP+HE                                                                       & LR                     & $\ge 2$                     & \multicolumn{1}{c|}{A}             & N         & \multicolumn{1}{c|}{$\checkmark$} & $\checkmark$ \\ \hline
\cite{tan2022residue}                                                               & DP+HE                                                                       & LR                     & 2                           & \multicolumn{1}{c|}{P}             & N         & \multicolumn{1}{c|}{}             & $\checkmark$ \\ \hline
FederBoost\cite{tian2020federboost}                                                 & DP                                                                          & Tree                   & $\ge 2$                     & \multicolumn{1}{c|}{P}             & Y         & \multicolumn{1}{c|}{$\checkmark$} &              \\ \hline
OpBoost\cite{li2022opboost}                                                         & DP                                                                          & Tree                   & $\ge 2$                     & \multicolumn{1}{c|}{P}             & N         & \multicolumn{1}{c|}{$\checkmark$} &              \\ \hline
PIVODL\cite{zhu2021pivodl}                                                          & DP+HE                                                                       & Tree                   & $\ge 2$                     & \multicolumn{1}{c|}{P}             & N         & \multicolumn{1}{c|}{$\checkmark$} & $\checkmark$ \\ \hline
Grafting-LDP\cite{takahashi2023eliminating}                                         & DP                                                                          & Tree                   & $\ge 2$                     & \multicolumn{1}{c|}{P}             & N         & \multicolumn{1}{c|}{}             & $\checkmark$ \\ \hline
TPSL\cite{yang2022differentially}                                                   & DP                                                                          & NN                     & 2                           & \multicolumn{1}{c|}{P}             & N         & \multicolumn{1}{c|}{}             & $\checkmark$ \\ \hline
\cite{ranbaduge2022differentially}                                                  & DP                                                                          & NN                     & $\ge 2$                     & \multicolumn{1}{c|}{P}             & N         & \multicolumn{1}{c|}{$\checkmark$} &              \\ \hline
VFGNN\cite{chen2020vertically}                                                      & DP                                                                          & NN                     & $\ge 2$                     & \multicolumn{1}{c|}{P}             & N         & \multicolumn{1}{c|}{$\checkmark$} & $\checkmark$ \\ \hline
DP-CutMixSL\cite{oh2022differentially}                                              & DP                                                                          & NN                     & $\ge 2$                     & \multicolumn{1}{c|}{P}             & N         & \multicolumn{1}{c|}{$\checkmark$} & $\checkmark$ \\ \hline
R3eLU\cite{mao2022secure}                                                           & DP                                                                          & NN                     & 2                           & \multicolumn{1}{c|}{P}             & N         & \multicolumn{1}{c|}{$\checkmark$} &              \\ \hline
PBM-VFL\cite{tran2023privacy}                                                           & DP+MPC                                                                          & NN                     & $\ge 2$                           & \multicolumn{1}{c|}{P}             & N         & \multicolumn{1}{c|}{$\checkmark$} &              \\ \hline
\end{tabular}
\end{table}

\section{VFL Phase: Model Deployment}
\label{sec:modeldeployment}
It has been shown that an adversary can exploit the trained model to infer private data~\cite{fu2022label,yin2020dreaming}, without needing access to the gradients and intermediate results from training and inference processes. These types of attacks are categorized as occurring during the model deployment phase, a stage where the model is fully trained and accessible to the adversary.

\subsection{Label Inference Attack}
\noindent\textit{Model Completion Attack.}
With the knowledge of the passive party model, the adversary requires auxiliary labeled data to train a complete model for label inference.
Fu et al. \cite{fu2022label} proposed a model completion attack that uses auxiliary labeled examples to train a complete model that is formed by
adding prediction layers to the passive party's model. The adversary can use the fine-tuned complete model to predict the labels of all training data or any new data. To improve the effectiveness of the attack, the authors further proposed an active model completion attack with malicious local optimizer that makes the top model in the active party more dependent on the bottom model of the malicious passive party via scaling gradients in the stochastic gradient descent training process. In this way, the attacker can complete the model with better inference accuracy of the labels.

\subsection{Feature Inference Attack}
Recent research has shown that attackers may recover training data solely from the parameters of trained models. Yin et al.~\cite{yin2020dreaming} proposed to exploit the parameters of the batch normalization layers to recover training data, as these layers store running means and variances of the activations of training data in multiple layers. However, it is worth noting that there are no existing works in VFL that we have come across which exploit the parameters of trained models at the active party to reconstruct a passive party's data. 

\subsection{Model Extraction Attack}
\noindent\textit{Passive Party Steals from Active Party.}
For attackers with access to the passive party's model, they can derive an accurate surrogate of the active party's model through supervised learning with a subset of training data or auxiliary labeled data. 
Model Completion Attack~\cite{fu2022label} used for label inference can be viewed as an attack that extracts the active party's model. Li et al.~\cite{li2023model} proposed the Train-ME attack, which uses the same strategy and obtains a surrogate of the active party's model. Neither of these attacks need access to any gradients during training nor do they make any inference query to the active party, but they require auxiliary labeled data.

\noindent\textit{Active Party Steals from Passive Party.}
It is yet unknown, if not impossible, how to extract models of passive parties by the active party without any information exchange. 

\subsection{Defenses}
Effective defenses can be implemented during training to defend against inference attacks during the model deployment phase. The basic idea is to make the trained local model unusable for effective model completion or accurate feature inference. 
\vspace{5pt}
\subsubsection{Non-Cryptographic Defense}~

\noindent\textit{Adversarial Training Defense against Model Completion Attacks:}
The goal of adversarial training strategies here is to hinder model completion attacks from attaining high label accuracy.
To defend the label inference attack with model completion in split learning, Zheng et al.\cite{zheng2022making} proposed an approach that pushes data points to their decision boundary by adding a potential energy loss to the loss function during training time. Potential energy loss characterizes the repulsive force between each pair of same-class outputs from the local model at the passive party. It aims to increase the generalization error when the attacker fine-tunes the complete model with labeled examples. Sun et al. \cite{sun2023robust} introduced a defense approach named DIMIP to protect the active party's labels from model completion attacks. DIMIP mitigates label leakage by adding an additional training objective that minimizes the mutual information between the passive party's representation and the true label. These defenses against the model completion attack not only mitigate the label leakage, but also make it difficult to obtain accurate surrogate models in model extraction attacks.

\vspace{5pt}
\subsubsection{Cryptographic Defense.} It's important to highlight that the previously mentioned model completion attacks rely on the assumption of having read access to the plaintext model weights. Consequently, if the training process is conducted using HE or MPC protocols with the model being encrypted, and the attacking party lacks the decryption key for their local models, they are unable to leverage their local models to implement these types of attack. This can be achieved by letting the active party or trusted coordinator handle the HE key pair distribution and gradient computation under HE during training. The passive party cannot decrypt its local model for model completion.

\begin{table}[]
\scriptsize
\setlength{\tabcolsep}{1pt}
\caption{Summary of existing attacks and defenses in model deployment phase of VFL.}\label{tab:attackdefenseindeploy}
\begin{tabular}{|c|c|c|c|c|c|c|c|c|c|c|c|}
\hline
Attacks                                                                                                   & \begin{tabular}[c]{@{}c@{}}Attack \\ Strategy\end{tabular} & Goal                                               & Knowl & \begin{tabular}[c]{@{}c@{}}Attack\\ Direction\end{tabular} & \begin{tabular}[c]{@{}c@{}}Attack \\ Type\end{tabular} & Attacker & \# parties& \begin{tabular}[c]{@{}c@{}}Model\\ (\# classes)\end{tabular} & \begin{tabular}[c]{@{}c@{}}HE \\ proto\end{tabular} & Defenses                                              & \begin{tabular}[c]{@{}c@{}}Defense \\ Strategy\end{tabular} \\ \hline
Model Completion\cite{fu2022label}                                                                        & S-S                                                        & \begin{tabular}[c]{@{}c@{}}LIA/\\ MEA\end{tabular} & K-A   & P->A                                                       & B-A                                                    & P       & $\ge 2$ & \begin{tabular}[c]{@{}c@{}}NN \\ (>=2)\end{tabular}          & N                                                      & \multirow{3}{*}{\cite{zheng2022making,sun2023robust}} & \multirow{3}{*}{AT}                                         \\ \cline{1-10}
\begin{tabular}[c]{@{}c@{}}Active Model Completion\\ with Gradient Scaling\cite{fu2022label}\end{tabular} & S-S                                                        & \begin{tabular}[c]{@{}c@{}}LIA/\\ MEA\end{tabular} & K-A   & P->A                                                       & B-A                                                    & A       & $\ge 2$ & \begin{tabular}[c]{@{}c@{}}NN \\ (>=2)\end{tabular}          & N                                                      &                                                       &                                                             \\ \cline{1-10}
Train-ME\cite{li2023model}                                                                                & S-S                                                        & \begin{tabular}[c]{@{}c@{}}LIA/\\ MEA\end{tabular} & K-A   & P->A                                                       & B-A                                                    & P       & $\ge 2$ & \begin{tabular}[c]{@{}c@{}}NN \\ (>=2)\end{tabular}          & N                                                      &                                                       &                                                             \\ \hline
\end{tabular}
\end{table}

Table \ref{tab:attackdefenseindeploy} summarizes the attacks and defenses discussed in the model deployment phase. The `Goal' column indicates that these attacks can be used for both label inference and model extraction attack. Existing defenses, specifically designed to counter these attacks, are primarily based on adversarial training.

\section{VFL Phase: Model Inference}\label{sec:modelinference}
In this section, we examine privacy threats in the inference stage of the model, which aim to extract information about the samples used for inference and models.

\subsection{Label Inference Attack}
The \textit{Spectrum-based} Attack~\cite{sun2022label} discussed in the model training stage can also be used during the model inference time. By participating in the inference on sufficient samples, a passive party can predict their labels by clustering the output of its trained local model, with only auxiliary information on the population-level information about the distributions of the positive and negative class's cut-layer embedding. 

\subsection{Feature Inference Attack}
In the feature inference attack, an adversary uses prediction results or intermediate results (e.g., forwarding embeddings) or both to reconstruct a sample or partial features of a sample. Unless otherwise noted, the active party is the adversary, aiming to infer the private features of the passive parties. We categorize existing attacks into two types based on the information they use: prediction-based and intermediate result-based attacks.

\subsubsection{Prediction Based Attacks} This type of attack is based on the prediction output (often provided as a confidence score) to recover the private feature input.

\textit{1. Equality solving attack.}
Under the assumption that the active party is the attacker and knows all the parameters of the model, as well as part of the features of a sample, Luo et al. \cite{luo2021feature} proposed an equality solving attack against the logistic regression (LR) model. When the prediction output (i.e., confidence score) of a sample is obtained, the adversary can construct a system of equations based on this prediction, and solve these equations to obtain the unknown features.

\textit{2. Path restriction attack.}
Luo et al. \cite{luo2021feature} also introduced a path restriction attack targeting the decision tree model.
Using partial feature values, the adversary can limit the possible paths leading to the predicted class. By doing so, they can estimate the range of unknown features. 

\textit{3. Generative model based Attacks.}
Besides direct attacks from an individual prediction discussed above, Luo et al. \cite{luo2021feature} also proposed an attack against complex models such as DNN based on multiple prediction outputs accumulated by the adversary during inference time. It trains a Generative Regression Network to predict unknown target features by minimizing the loss between the ground-truth predictions and the predictions made for samples with unknown features treated as random variables, along with known feature values.

Yang et al. \cite{yang2023practical}proposed a feature inference attack in VFL applications for Internet of Things(IoT). They focused on a more practical black-box situation where there are two passive parties and one active party. The adversary is a passive party with only a bottom model and aims to infer private features at another passive party. The attack assumes that the adversary can query the VFL service and feed any input to the model to obtain the prediction score. Given that, the adversary can train a generative model to predict private features of the target passive party by minimizing the distance between the real confidence scores during inference collected on real samples and the confidence scores queried by the adversary with the estimated features. Because the attacker cannot access the top model and other bottom models to obtain the gradients of the parameters, the zeroth-order gradient estimation method is used to train the generative model.

\textit{4. Model inversion attacks.}
Jiang et al. \cite{jiang2022comprehensive} proposed two model inversion attacks in white-box and black-box settings, respectively. For the white-box attack GIAUW, the active party has full access to the entire model, including the model component at the passive parties.  Given the inference result of a sample, i.e., the confidence score $c$, the adversary iteratively optimizes the estimated passive party input by minimizing the distance between $c$ and the confidence score deduced from the estimated input.  For the black-box attack GIAUB, the adversary does not know the model component at the passive parties but has access to auxiliary data which is a small set of passive-party data. The adversary first uses the auxiliary data to train a shadow model as a surrogate to the passive party's model, and then applies the white-box attack to infer the input feature. Compared to previous works \cite{luo2021feature}, attacks errors can be reduced by two to three times.

\subsubsection{Intermediate Result based Attacks} This type of attack uses the intermediate result forwarded from passive parties to recover their private data, directly or indirectly. We discuss existing attacks within three categories, as given below: equation solving attacks, adversarial training-based attacks, and model extraction-based attacks. The equation-solving attacks directly compute the input private features by solving a system of equations.  The adversarial training-based and model extraction-based attacks are indirect attacks, because they compute a model during the training time, which is then used to derive the private input features during the inference time.

\textit{1. Equation solving attacks.}
We discussed an intermediate data-based attack proposed by Ye et al. \cite{ye2022feature} in Section \ref{ssec:resdataintraining}. Because the attack only requires the intermediate results to reconstruct the private binary input features of the passive party, it can also be applied to the model inference time.

\textit{2. Generative model based attacks.}
In Section \ref{ssec:resdataintraining}, we discussed a feature space hijacking attack (FSHA)\cite{pasquini2021unleashing} that trains an adversarial network to recover the input features of passive parties from its smashed data output (i.e., intermediate results). The attack is setup during the VFL training process. Once the setup is completed and the GAN model is trained, the adversary can use it to recover instances in the later stage of the training process, as well as in the entire VFL inference stage. Gawron et al.~\cite{gawron2022feature} applied FSHA to the learning process of a split neural network enhanced with differential privacy (DP) using a client-side off-the-shelf DP optimizer. Their result shows that FSHA is robust and that private data can still be successfully reconstructed with low error rates at arbitrary DP $\epsilon$ levels as long as the differentially private client model achieves sufficient accuracy.

\textit{3. Model extraction based attacks.}
Model extraction attacks can be exploited by the active party to infer the private features of the passive party. 
Gao et al.~\cite{gaopcat} developed the PCAT attack for model extraction (as discussed in \ref{ssec:acpameintraining}), and further proposed a data reconstruction attack based on it. After extracting the model from the passive party, the adversary uses auxiliary labeled samples to train a reverse mapping $f^{-1}$ to reconstruct the input of the extracted model from the corresponding output. During inference time, the adversary feeds the intermediate result received from the passive party into $f^{-1}$ to reconstruct its input. The adversary can further fine-tune the reconstructed data by minimizing the difference between the ground-truth intermediate result and the output of the extracted model.

In addition to label inference and feature inference attacks, Qiu et al \cite{qiu2022your} also proposed a Label-related Relation Inference (LRI) attack targeting edges in graph data (i.e. connections between samples) where the labels are strong indicators of the existence of such connections. In LRI attack, the adversary is a passive party and has access to the global module to obtain prediction scores. The target party is a passive party that holds the relation information for training. The adversary first recovers the model output of the target party by iterative optimization to minimize the distance between the ground-truth scores and the scores produced with its current estimate. The relation between two samples can be determined by computing the similarity between the corresponding model outputs from the target.

\subsection{Model Extraction Attack}
\subsubsection{Passive Party Steals from Active Party}
The model extraction attack discussed during model deployment can also occur during the model inference stage because the passive party can still derive a surrogate of the active party's model as long as it has auxiliary labeled data.  Without the auxiliary data, an adversary can resort to a data-free attack to extract models by querying the active party's model prediction APIs in the model inference stage. For example, Truong et al. \cite{truong2021data} proposed a GAN-based data-free model extraction attack, which trains a generator to generate fake inputs into the active party model $V$ and a surrogate model $S$, and uses confidence score matching to train $S$ with the objective of minimizing disagreement between $V$ and $S$. An adversary in the passive party can use this attack to obtain $S$ as a surrogate of the active party's model. 

\subsubsection{Active Party Steals from Passive Party}
Besides the training-time PCAT attack discussed in Section \ref{ssec:meintraining}, Gao et al.~\cite{gaopcat} also introduced a vanilla attack where the server aims to steal the client's model (i.e., the passive party's model) after training. It assumes that the server does not know anything about the client model structure but collects a limited number of training samples from the same domain. By using these auxiliary data during the inference time, it trains a model as a surrogate to the client's model by minimizing the KL divergence between the output soft labels from the real VFL model and the output produced with the surrogate model.     
The active party could also employ the Unsplit attack~\cite{erdougan2022unsplit} discussed in Section \ref{ssec:acpameintraining} which trains a surrogate model mirroring the passive party's model by minimizing the distance between the outputs of the two models, as long as the active party can initiate queries on a set of fixed examples.

Table \ref{tab:attackinference} summarizes all the attacks that we reviewed during the model inference phase. It is worth noting that, although not originally proposed for MEA during the VFL inference phase, both Datafree* and Unsplit* attacks can still be applicable in this scenario.

\begin{table}[]
\scriptsize
\setlength{\tabcolsep}{1pt}
\caption{Summary of existing attacks in model inference phase of VFL.}\label{tab:attackinference}
\begin{tabular}{|c|c|c|c|c|c|c|c|c|c|c|}
\hline
Attack                & \begin{tabular}[c]{@{}c@{}}Existing \\ Attacks\end{tabular} & \begin{tabular}[c]{@{}c@{}}Attack \\ Strategy\end{tabular} & Method                                                                               & Knowledge     & \begin{tabular}[c]{@{}c@{}}Attack \\ Direction\end{tabular} & \begin{tabular}[c]{@{}c@{}}Attacker \\ Type\end{tabular} & \begin{tabular}[c]{@{}c@{}}Attacker \\ (Capability)\end{tabular} &\# parties & \begin{tabular}[c]{@{}c@{}}Model       \\ (\# classes)\end{tabular} & \begin{tabular}[c]{@{}c@{}}HE \\ proto\end{tabular} \\ \hline
LIA                   & \cite{sun2022label}                                         & I-S                                                        & Spectrum Attack                                                                      & K-I, K-C      & P->A                                                        & B-A                                                      & P             &            2                                       & NN (2)                                                             & N                                                      \\ \hline
\multirow{10}{*}{FIA} & ESA\cite{luo2021feature}                                    & P-S                                                        & Equality solving                                                                     & K-S, K-M      & A->P                                                        & W-A                                                      & P              &           $\ge 2$                                       & LR($\ge 2$)                                                        & N                                                      \\ \cline{2-10} 
                      & PRA\cite{luo2021feature}                                    & P-S                                                        & Path restriction                                                                     & K-S, K-M      & A->P                                                        & W-A                                                      & P                       &               $\ge 2$                          & Tree($\ge 2$)                                                      & N                                                      \\ \cline{2-11} 
                      & GRNA\cite{luo2021feature}                                    & P-S                                                        & Generative regression                                                                & K-S, K-M      & A->P                                                        & W-A                                                      & P                        &             $\ge 2$                           & NN,Tree ($\ge 2$)                                                  & N                                                      \\ \cline{2-11} 
                      & \cite{yang2023practical}                                    & P-S                                                        & Generative model                                                                     & K-S           & P->P                                                        & B-A                                                      & P                          &             3                        & NN ($\ge 2$)                                                       & N                                                      \\ \cline{2-11} 
                      & GIAUW\cite{jiang2022comprehensive}                          & P-S                                                        & Gradient Inversion                                                                   & K-S, K-M      & A->P                                                        & W-A                                                      & P                          &              2                        & LR,NN($\ge 2$)                                                     & N                                                      \\ \cline{2-11} 
                      & GIAUB\cite{jiang2022comprehensive}                          & P-S                                                        & \begin{tabular}[c]{@{}c@{}}Surrogate model\\ +Gradient Inversion\end{tabular}        & K-S, K-A      & A->P                                                        & B-A                                                      & P                           &              2                       & LR,NN($\ge 2$)                                                     & N                                                      \\ \cline{2-11} 
                      & \cite{ye2022feature}                                        & I-S                                                        & Equation Solving                                                                     & K-I           & A->P                                                        & B-A                                                      & P                             &             2                      & NN (>=2)                                                           & N                                                      \\ \cline{2-11} 
                      & FSHA\cite{pasquini2021unleashing}                           & I-S                                                        & GAN+Adversarial Training                                                             & K-I, K-A      & A->P                                                        & B-A                                                      & A(L-M)                        &            2                       & NN (>=2)                                                           & N                                                      \\ \cline{2-11} 
                      & \cite{gaopcat}                                              & I-S                                                        & \begin{tabular}[c]{@{}c@{}}PCAT Model Extraction \\ + Model Inversion\end{tabular}   & K-S, K-A      & A->P                                                        & B-A                                                      & P                              &           $\ge 2$                       & NN (>=2)                                                           & N                                                      \\ \cline{2-11} 
                      & \cite{qiu2022your}                                          & P-S                                                        & \begin{tabular}[c]{@{}c@{}}Representation inversion\\ +similarity\end{tabular}       & K-S, K-M      & P->P                                                        & G-A                                                      & P                               &            $\ge 2$                     & GNN(>=2)                                                           & Y                                                      \\ \hline
\multirow{3}{*}{MEA}  & Datafree*\cite{truong2021data}                           & S-S                                                        & \begin{tabular}[c]{@{}c@{}}Confidence score matching\\ for GAN training\end{tabular} & K-S           & P->A                                                        & B-A                                                      & P                                  &               2               & NN (>=2)                                                           & N                                                      \\ \cline{2-10} 
                      & Unsplit*\cite{erdougan2022unsplit}                          & S-S                                                        & \begin{tabular}[c]{@{}c@{}}Intermediate result \\ divergence loss\end{tabular}       & K-I, K-H, K-A & A->P                                                        & G-A                                                      & P         &                     2                                  & NN (>=2)                                                           & N                                                      \\ \cline{2-11} 
                      & Vanilla\cite{gaopcat}                                       & S-S                                                        & \begin{tabular}[c]{@{}c@{}}soft-label prediction \\ divergence loss\end{tabular}     & K-S, K-A      & A->P                                                        & B-A                                                      & P          &                  $\ge 2$                                    & NN (>=2)                                                           & N                                                      \\ \hline
\end{tabular}
\end{table}

\subsection{Defenses}
\subsubsection{Non-Cryptographic Defense}~\\
\noindent\textit{Prediction Obfuscation:}
To defend the prediction-based inference attack, a straightforward solution is to obscure the prediction output. 
Luo et al. \cite{luo2021feature} discussed the countermeasure that rounds the confidence score to less precision to reduce the effectiveness of the equality solving attack. The authors also discussed other potential countermeasures such as dropout for neural networks, feature preprocessing, and output privacy verification. In addition to rounding defense, Jiang et al.~\cite{jiang2022comprehensive} also considered two other different defense methods to obscure confidence scores: noising defense and purification defense. The noising defense simply perturbs the confidence
scores with the random noise sampled from a Gaussian distribution. The purification defense uses a pre-trained autoencoder that is trained on real confidence scores to produce confidence scores with less dispersion for the inputs in the same class.

\noindent\textit{Feature Obfuscation:}
Luo et al. \cite{luo2021feature} discussed the potential of feature preprocessing as a means to mitigate the feature privacy leakage. They found that ESA,PRA and GRNA attacks tend to be more effective when the number of classes is relatively large or the target's features are highly correlated with the adversary's known features. Thus, they suggest that removing highly related features from the inference input could be an effective defense strategy against these attacks.   

\subsubsection{Cryptographic Defense.} 
It should be noted that cryptographic security in VFL inference is typically achieved through cryptographic defense strategies that are deployed during the VFL training phase and implement secure protocols for inference over encrypted messages. Our prior discussion reveals that the spectrum-based label inference attack presupposes plaintext access to the local model's output. Similarly, prediction-based feature inference attacks require access to the plaintext of the final prediction output, and intermediate-result-based feature inference attacks rely on plaintext access to the intermediate results outputted by passive parties. If HE and MPC based defenses during training ensure that the input and output of local model at the attacking party are encrypted and cannot be decrypted by the attacker, they can serve effective measures to thwart these attacks.

To the best of our knowledge, there have been no inferenece-time only solutions specifically for secure inference in the context of VFL. However, systems like Cheetah~\cite{huang2022cheetah} exemplify secure model inference systems that utilize cryptographic protocols specifically for the inference computation of general machine learning models. They involve a set of cryptographic protocols for the common linear operations and nonlinear operations of DNNs. After the training phase of the VFL model, participants in the VFL can use these systems to defend against privacy attacks during inference. This is achieved by mandating that any party, including the insider attacker, processes encrypted inputs in accordance with the established cryptographic protocols for inference and guaranteeing that the encryption keys are inaccessible to the attacker. Such measures can effectively protect the model input and output.

\begin{table}[]
\scriptsize
\setlength{\tabcolsep}{1pt}
\caption{Summary of existing defenses in model inference phase of VFL.}\label{tab:defenseinference}
\begin{tabular}{|c|c|c|c|cc|cc|}
\hline
\multirow{2}{*}{\begin{tabular}[c]{@{}c@{}}Non-Crypto\\ Defenses\end{tabular}} & \multirow{2}{*}{\begin{tabular}[c]{@{}c@{}}Defense\\ Strategy\end{tabular}} & \multirow{2}{*}{Model} & \multirow{2}{*}{\# parties} & \multicolumn{2}{c|}{Threat Model}              & \multicolumn{2}{c|}{Privacy}              \\ \cline{5-8} 
                                                                               &                                                                             &                        &                            & \multicolumn{1}{c|}{Attacker type} & collusion & \multicolumn{1}{c|}{feature}      & label \\ \hline
Feature preprocessing\cite{luo2021feature}                                     & DO(input feature)                                                           & LR, Tree, NN           & $\ge 2$                    & \multicolumn{1}{c|}{P}             & N         & \multicolumn{1}{c|}{$\checkmark$} &       \\ \hline
Rounding score \cite{luo2021feature,   jiang2022comprehensive}                 & DO(prediction score)                                                        & LR, NN                 & $\ge 2$                    & \multicolumn{1}{c|}{P}             & N         & \multicolumn{1}{c|}{$\checkmark$} &       \\ \hline
Noisy score \cite{jiang2022comprehensive}                                      & DO(prediction score)                                                        & LR, NN                 & $\ge 2$                    & \multicolumn{1}{c|}{P}             & N         & \multicolumn{1}{c|}{$\checkmark$} &       \\ \hline
Score purification   \cite{jiang2022comprehensive}                             & DO(prediction score)                                                        & LR, NN                 & $\ge 2$                    & \multicolumn{1}{c|}{P}             & N         & \multicolumn{1}{c|}{$\checkmark$} &       \\ \hline
\end{tabular}
\end{table}

Table \ref{tab:defenseinference} summarizes the existing non-cryptographic defenses during the VFL inference phase. To the best of our knowledge, no existing work has been proposed to address the label inference attack during the inference phase.

\section{Summary, Open Challenges and Future Direction}
In this paper, we conduct a comprehensive review of the latest research on privacy attacks and defenses in VFL. Our survey is uniquely structured through the lens of the machine learning life cycle, providing insights that help researchers identify distinct challenges posed at different stages, and assist practitioners in implementing appropriate security solutions during the development. We provide a systematic and detailed taxonomy that characterizes privacy attacks and defense mechanisms in VFL, considering aspects like threat models, capabilities, technical strategies, and application scenarios.

The field of VFL privacy is rapidly evolving, presenting numerous challenges and opportunities. In the following, we identify existing gaps in current research on VFL privacy, discuss some open challenges, and suggest potential directions for future research. 

\subsection{Gaps in Privacy Threat Research}
\noindent\textbf{Privacy threats in multi-party VFL.}\quad As illustrated in Tables \ref{tab:attacksintraining}-\ref{tab:attackinference}, the majority of existing research on privacy attacks against VFL focus on two-party VFL. However, with VFL's evolving application in real-world scenarios, multi-party VFL, involving three or more participants for joint modeling, could become a popular setting. It is still not clear if the existing attacks can be easily and effectively extended to multi-party VFL. Moreover, multi-party VFL may introduce new vulnerabilities. In scenarios involving collusion, the adversary could gain more capabilities to infer private data. Privacy attacks can occur not only between the active party and passive parties but also among passive parties. Our review found only one study \cite{qiu2022your} that considers `P->P' attack direction as shown in Table \ref{tab:attackinference}. On the other hand, although many existing defense mechanisms, listed in Tables \ref{tab:cryptodefenseintraining}-\ref{tab:defenseinference}, are developed within multi-party VFL context, the absence of specific privacy attacks targeting this setting leaves questions about their robustness. Therefore, developing a comprehensive understanding of privacy risks in multi-party VFL is a crucial direction for future research.

\noindent\textbf{Privacy threats in VFL for tree models.}\quad  Comparing Tables \ref{tab:attacksintraining}, \ref{tab:attackdefenseindeploy}, \ref{tab:attackinference} with Table \ref{tab:cryptodefenseintraining}, our review reveals a notable disparity: there are significantly fewer studies on attacks against VFL tree models than defense works. Given the most defense mechanisms for tree models achieve S-1 and S-2 security levels, the lack of thorough research on potential privacy attacks in these scenarios leaves the robustness of current solutions unclear.
Therefore,  it is essential to focus on analyzing privacy threats and developing effective attacks against VFL tree models to fully assess and enhance their security.

\subsection{Gaps in Privacy Defense Research}
\noindent\textbf{Limitations of non-cryptographic defenses.} \quad
The current body of non-cryptographic defense research has two primary limitations: 1. Existing studies employing DO and AT do not account for adaptive adversaries who may alter their attack strategies in response to different defense mechanisms and their configuration 2. Most research does not consider the threat of collusion in a multi-party VFL context. These gaps result in a lack of comprehensive understanding of their robustness in practical scenarios. To overcome these challenges, it is crucial to develop robust defenses that are effective against adaptive and collusive adversary models.

\noindent\textbf{Limitations of cryptographic defenses.} \quad
As indicated in Table \ref{tab:cryptodefenseintraining}, many proposed defenses achieves S-1 security level, which is insufficient against attacks that exploit partial training information, such as aggregated gradients. Although the mechanisms at the S-2 and S-3 security levels appear more promising to defend against many training-time attacks, they significantly increase computation and communication costs. This creates a challenging trade-off between efficiency and privacy, which remains a critical focus in VFL research. A possible route to address this issue is that, through a detailed analysis of privacy threats specific to VFL applications, we develop hybrid mechanisms that combine different strategies and tailor them to different privacy requirements to improve efficiency while maintaining robustness against privacy attacks.   

\subsection{Gaps in End-to-End VFL Privacy}
\noindent\textbf{Privacy attacks and defenses in VFL deployment and inference phase.} \quad  Our review highlights that current research predominantly focuses on privacy attacks and defenses during the training phase of VFL, with considerably less attention to the deployment and inference stages. Yet, examining VFL through the lens of the machine learning lifecycle reveals that privacy risks persist even after a model is securely trained, particularly during deployment and inference.  Therefore, it is imperative to conduct a more in-depth examination of privacy threats and defense strategies during these phases. Topics such as restricting model access during deployment and strengthening protections during the inference stage should be further explored to effectively defend against privacy attacks.

\subsection{Future Directions}
\noindent\textbf{Understanding  unique characteristics of privacy in VFL.} \quad
In Section \ref{ssec:privacyvfl}, we list distinct privacy considerations in VFL in contrast to HFL. Our review reveals that current research focuses primarily on privacy risks stemming from the exchange of direct data representations, without considering the effect of three other aspects. The exchange frequency on the effectiveness of privacy attacks has not been thoroughly understood. Additionally, feature correlation in aligned samples could potentially exacerbate label and feature inference attacks, as well as model extraction attacks. Yet, these issues have received limited attention in research. A deeper understanding of these unique characteristics in VFL is crucial. It can lead to the identification of new vulnerabilities and the development of more effective and efficient defense mechanisms.

\noindent\textbf{Self-Supervised VFL for privacy.} \quad
Existing research on VFL privacy has predominantly focused on the training process with aligned samples. However, Self-supervised VFL, as discussed in recent studies by Castiglia et al.~\cite{castiglia2022selfsupervised} and Yang et al.~\cite{yang2022split}, presents an innovative approach that improves model performance using unaligned samples while also offering privacy benefits. In this approach, each passive party uses unsupervised learning on its own data to train a local model that converts training samples to embeddings. These embeddings are then used by the active party to train its top model. This methodology avoids the exchange of gradients, thereby mitigating the risk of gradient-based attacks and safeguarding label privacy of the active party. Therefore, a promising avenue for advancing VFL privacy involves the development of privacy-preserving self-supervised VFL strategies. These strategies should aim to maximize the utility of unaligned samples to improve performance while preventing any potential leakage from the embeddings.     

\noindent\textbf{Privacy Auditing for VFL.} \quad
With the development of defense mechanisms like differential privacy and other types of non-cryptogprahic defenses,  there's a critical need for a quantitative methodology to assess the practical effectiveness of these privacy protections. Existing auditing approaches for differentially private Stochastic Gradient Descent (SGD) ~\cite{NEURIPS2020_fc4ddc15} and HFL~\cite{li2022auditing,andrew2023one} cannot seamlessly extend to the unique characteristics of VFL. Consequently, the development of an auditing technique, specifically tailored for VFL, stands as a vital future direction.

In conclusion, while VFL presents an increasingly attractive solution for industries to solve data silo problems due to stringent data regulations, its privacy research is still in a nascent stage. It is our hope that this work serves as a catalyst for future research endeavors aimed at addressing these critical challenges in the field.

\clearpage
\bibliographystyle{unsrt}
\bibliography{sample-base}

\end{document}